\documentclass[aps, prd, twocolumn, superscriptaddress, floatfix]{revtex4-2}
\usepackage{graphicx}
\usepackage{amsmath}
\usepackage{amssymb}
\usepackage{tikz}
\usepackage{braket}
\usepackage{pgfplots}
\usepackage{slashed}
\usepackage{xfrac}
\usepackage[version=4]{mhchem}
\pgfplotsset{compat=1.17}
\usepackage{subcaption}
\usetikzlibrary{shapes.geometric}
\usetikzlibrary{shadows,patterns,perspective}
\usetikzlibrary {decorations,decorations.text}
\usetikzlibrary{decorations.pathreplacing}
\usepackage{float}
\usepackage{hyperref}
\usepackage{xcolor,ulem}
        
\definecolor{linkcolor}{RGB}{0, 0, 255}  
\definecolor{citecolor}{RGB}{0, 128, 0}   
\definecolor{urlcolor}{RGB}{255, 0, 0}    
\usepackage{pgfplots}
\pgfplotsset{compat=1.17}

\hypersetup{
    colorlinks=true,
    linkcolor=linkcolor,
    citecolor=citecolor,
    urlcolor=urlcolor,
    linktoc=all,  
    pdfborder={0 0 0}  
}

\usetikzlibrary{decorations.markings,calc,decorations.pathreplacing}

\DeclareCaptionJustification{justified}{\leftskip=0pt \rightskip=0pt \parfillskip=0pt plus 1fil}

\begin{document}
\definecolor{dy}{rgb}{0.9,0.9,0.4}
\definecolor{dr}{rgb}{0.95,0.65,0.55}
\definecolor{db}{rgb}{0.5,0.8,0.9}
\definecolor{dg}{rgb}{0.2,0.9,0.6}
\definecolor{BrickRed}{rgb}{0.8,0.3,0.3}
\definecolor{Navy}{rgb}{0.2,0.2,0.6}
\definecolor{DarkGreen}{rgb}{0.1,0.4,0.1}

\title{Competing color superconductivity and color-Kondo effect in quark matter}

\author{Pradip Kattel}
\email{pradip.kattel@rutgers.edu}
 
\author{Abay Zhakenov}

\author{Natan Andrei}
\affiliation{Department of Physics and Astronomy, Center for Materials Theory, Rutgers University,
Piscataway, New Jersey 08854, United States of America}

\begin{abstract}
The competition between bulk color superconductivity and the localized screening of a heavy quark impurity, analogous to the Kondo effect, leads to a rich spectrum of phenomena in dense quark matter. We investigate this competition at the edge of a superconducting quark bulk, where both the superconducting gap and the Kondo scale are dynamically generated in a tractable toy model. Utilizing the exact Bethe Ansatz method, we elucidate the resulting boundary physics. We identify distinct regimes characterized by either multi-particle Kondo screening or an unscreened local moment. Crucially, we also uncover a novel intermediate phase featuring impurity screening through a single-particle bound state formed within the superconducting gap. The toy model presented in this work highlights the complex interplay between dynamically generated bulk properties and boundary impurities in extreme QCD environments, offering potential insights into phenomena occurring in heavy-ion collisions and compact stars.

\end{abstract}

\maketitle

\section{Introduction}

Quarks and gluons, as described by Quantum Chromodynamics (QCD) \cite{aitchison2012gauge,bailin1993introduction,marciano1978quantum,greiner2007quantum}, are typically strongly interacting and confined within protons and neutrons. They become weakly coupled (asymptotically free)  under extremely high temperatures or densities, undergoing deconfinement and forming a Quark-Gluon Plasma (QGP). This occurs at temperatures above the pseudo-critical value $T_c \approx 155~\mathrm{MeV}$, as established by lattice QCD~\cite{Borsanyi2014,Bazavov2014} and confirmed in heavy-ion collisions at RHIC and the LHC~\cite{Yagi2005,Sarkar2009,Letessier2002,Pasechnik2017,Berges2021}. At high baryon density ($\mu_B \gtrsim 500~\mathrm{MeV}$), deconfinement is instead expected to occur at significantly lower temperatures, in the range $T \sim 50\text{--}100~\mathrm{MeV}$, depending on the location of the critical point and the structure of the QCD phase diagram~\cite{Kurkela2016}. Below this range, QCD is expected to host a wealth of exotic phases such as color superconductivity and quarkyonic matter~\cite{Rajagopal2001,Huang2005,McLerran2007}, which are actively studied in the context of compact stars and dense matter experiments. The study of the phenomenologically relevant low-temperature regions, where novel phases emerge, has proved to be challenging due to strong coupling and inherent non-perturbative effects, making theoretical and experimental investigations particularly demanding~\cite{mclerran2019quarkyonic,aarts2016introductory}.

\begin{figure}
    \centering
    \includegraphics[width=\linewidth]{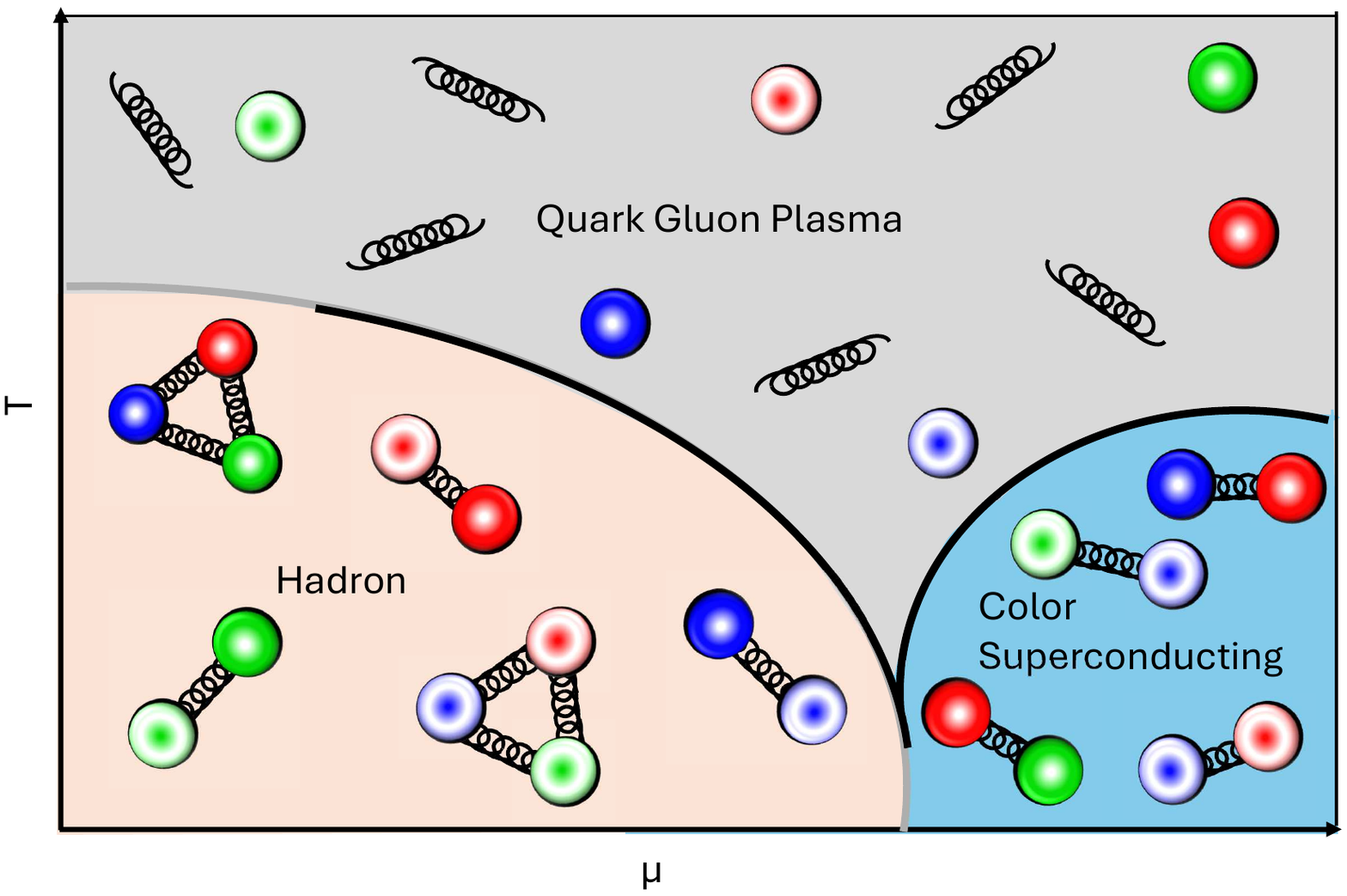}
    \caption{Phase diagram of QCD in temperature and chemical potential phase showing the hadronic, quark-gluon plasma, and color superconducting phase.}
    \label{fig:qcdphasediagram}
\end{figure}

One particularly compelling phase is color superconductivity~\cite{Alford:2001dt,alford2008color,Ren2003ColorSI}. In this phase, quarks near the Fermi surface form Cooper pairs through attractive interactions in color channels, leading to a finite energy gap in the excitation spectrum and a spontaneous breaking of color gauge symmetry~\cite{Shovkovy2004TwoLO} as shown in the phase diagram~\ref{fig:qcdphasediagram}. This state is believed to play a crucial role in the inner cores of neutron stars~\cite{Alford2002CompactSW,Alford2000ColorSI} and may be accessible in ultra-relativistic heavy-ion collisions~\cite{tomasik2006ultrarelativistic}, offering a window into the behavior of strongly interacting matter under extreme conditions. Color-superconducting quark matter has multiple phases because quarks, which come in three colors and six flavors (up, down, strange, charm, bottom, and top), can form Cooper pairs in various ways. Each pairing pattern breaks different symmetries, leading to distinct excitation spectra and transport properties. Predicting the favored pairing mechanism among the many possibilities is extremely challenging. In principle, a full QCD calculation in the strong coupling regime should determine the answer, but current lattice QCD methods have not yet achieved this. To gain quantitative insights, one often turns to the Nambu–Jona–Lasinio (NJL) model, a simplified version of QCD that captures key aspects of color superconductivity~\cite{Buballa:2003qv,gholami2025renormalization}. In its traditional formulation, the NJL model replaces dynamical gluons with an effective four-fermion interaction, which is not renormalizable in $3+1$ dimensions; hence, a UV cutoff $\Lambda$ must be introduced to define the theory as an effective model valid below that scale~\cite{de2015cosmology}. More recently, functional-renormalization-group (FRG) approaches have been employed to overcome the associated cutoff artefacts, yielding regulator-independent results in medium-dependent situations relevant for color superconductivity~\cite{gholami2025renormalizing}.

In parallel, recent theoretical studies have uncovered the possibility of a QCD analog of the Kondo effect~\cite{Hattori:2015hka,Yasui:2016svc,Fariello:2019ovo,adhikari2024strongly}. In this scenario, heavy quarks such as charm or bottom act as impurities within a sea of light quarks (up, down, and strange)~\cite{Suenaga:2019car}. The non-Abelian gluon exchange between light and heavy quarks generates an interaction reminiscent of the traditional Kondo effect observed in condensed matter systems, where conduction electrons screen magnetic impurities~\cite{kondo2012physics,hewson1993kondo,andrei1983solution,tsvelick1983exact,affleck1991kondo}. In quark matter, this interaction leads to the formation of a Kondo gap and the emergence of mixed states, where light and heavy quarks combine into Bogoliubov quasiparticles~\cite{Yasui:2016svc,yasui2017topology}. {
A related secondary Kondo mechanism between gapped and ungapped quarks in the two--flavor 
color--superconducting (2SC) phase was proposed in Ref.~\cite{hattori2019}, 
where the gapped quarks act effectively as heavy impurities interacting with ungapped modes.}

At high baryon density (or large quark chemical potential $ \mu $), quarks near the Fermi surface experience attractive interactions via gluon exchange, leading to Cooper pairing and the onset of color superconductivity, a non-
Abelian analog of BCS theory. The resulting superconducting gap behaves as
\begin{equation}
    \Delta \sim \Lambda \exp\left(-\frac{c}{G}\right),
\end{equation}
where $G$ is the 4-fermion attractive coupling, $c$ is a numerical constant determined by the pairing channel at the cut-off scale $\Lambda$.

In contrast, the presence of a heavy quark impurity allows for color exchange with the Fermi sea, resulting in logarithmic enhancement of the scattering amplitude at low energy—a hallmark of the QCD Kondo effect. The associated Kondo scale is related to cut-off scale $\Lambda$ by
\begin{equation}
    \Lambda_K \sim \Lambda \exp\left(-\frac{\pi}{2 N_c J}\right),\quad N_c=3\,,
\end{equation}
analogous to the Kondo temperature in condensed matter systems, where $J$ is the Kondo coupling constant.

Both phenomena occur near the Fermi surface, but compete with each other: superconductivity gaps out low-energy states, suppressing Kondo screening, while a developed Kondo cloud inhibits coherent pairing. The dominant ground state depends sensitively on the relative scale of $ \Delta $ and $ \Lambda_K $. 

To describe these phenomena in more detail, we proceed to discuss the theory in the regime of high quark chemical potential $ \mu \gg \Lambda_{\text{QCD}} $, where the QCD coupling becomes weak due to asymptotic freedom. In this regime, gluon-mediated interactions between quarks can be approximated by local four-fermion interactions, allowing an effective description in terms of an NJL-type model.

A model capturing both effects can be obtained starting from the standard QCD Lagrangian \footnote{
{

The QCD Lagrangian is
\begin{equation}
    \mathcal{L}_{\text{QCD}} = -\frac{1}{4} F^a_{\mu\nu} F^{a\mu\nu} + \bar{\psi}_i \left( i \gamma^\mu D_\mu - m_i \right) \psi_i 
\end{equation}
where the field strength tensor $ F^a_{\mu\nu} $ is defined as $F^a_{\mu\nu} = \partial_\mu A^a_\nu - \partial_\nu A^a_\mu + g f^{abc} A^b_\mu A^c_\nu,$
and the covariant derivative $ D_\mu $ is $D_\mu = \partial_\mu - i g \tau^a A^a_\mu$, with  $ A^a_\mu $ being the gluon fields, $ f^{abc} $  the structure constants of the SU(3) gauge group, $ \psi_i $ the quark fields with flavor index $ i $,  $ m_i $ is the mass of the $i$-th quark. and  $ \tau^a $ are the generators of the SU(3) group that acts on the 3-dimensional color space of quarks.}}, and  integrating out the gluon field, one finds the effective action in the quark sector \cite{kohyama2016possible}  
\begin{align}
    \mathcal{S}_{\mathrm{q}}&=\int \mathrm{d}x^4\bar\psi_i (i\slashed\partial -m)\psi_i\\
    &+G\int \mathrm{d}x^4\mathrm{d}y^4(\bar\psi_i\gamma^\mu \tau^a \psi_i)(x)D_{\mu\nu}^{ab}(x,y;\epsilon)(\bar\psi_j\gamma_\nu \tau^b \psi_j)(y) \nonumber
\end{align}
where $D(x,y;\epsilon)$ is the gluon propagator at some specific energy scale. In this limit, we shall now consider a single heavy quark $Q$ localized at $x=0$ and interacting with the lighter quarks $\psi_i$ via color-color interaction as
\begin{align}
    \mathcal{S}&_{\rm QCD-Kondo}=\mathcal{S}_{\rm q}\nonumber\\
    &+J\int \mathrm{d}x^4\mathrm{d}y^4(\bar\psi_i\gamma^0 \tau^a \psi_i)(x)D_{00}^{ab}(x,y;\mu) T^b\delta(y),
    \label{efflag}
\end{align}
where $T^b$ are the generators of $SU(3)$ acting on the color space of a localized heavy quark situated at $x=0$. We will consider here the impurity in the fundamental $SU(3)$ representation and defer to later publications other impurity representations.

{
Unlike electronic systems, where the impurity carries a spin-$\tfrac{1}{2}$ moment, 
the localized impurity here transforms in the fundamental color representation 
$\mathbf{3}$ of $\mathrm{SU}(3)$ and couples to the color current of the bulk quarks 
through the exchange term $J\,\psi_{+}^{\dagger}\tau^{a}\psi_{-}\,T^{a}$. 
When a bulk quark (in $\mathbf{3}$) interacts with the impurity (also in $\mathbf{3}$), 
the relevant tensor product decomposes as 
$\mathbf{3}\otimes\bar{\mathbf{3}}=\mathbf{1}\oplus\mathbf{8}$. 
The color-singlet component ($\mathbf{1}$) represents the screened configuration, 
analogous to the spin singlet in the $\mathrm{SU}(2)$ Kondo problem, 
while the octet ($\mathbf{8}$) corresponds to residual color excitations that remain gapped. 
Thus, the Kondo screening process in this model amounts to forming a many-body color-singlet between the impurity and the bulk, and the competition between this screening and 
the bulk pairing gap underlies the distinct impurity phases described in this work. While the overall structure of the phase diagram resembles that of the $\mathrm{SU}(2)$ case studied in Ref.~\cite{kattel2025thermodynamics}, the non-Abelian gapped excitation spectrum in both the bulk and boundary is considerably richer. For example, as we shall show later, within a portion of the Yu–Shiba–Rusinov (YSR) regime, the boundary spectrum contains an unexpected antiquark-like excitation, reflecting the nontrivial color structure of the screened impurity and the extended multiplet content of the $\mathrm{SU}(3)$ theory.}

In Ref.~\cite{kimura2019conformal}, the problem is studied in the limit $g=0$, where an S-wave approximation can be performed to write the effective theory in 1+1D 
as
\begin{equation}
    S^{2D}_{\text{eff}} = \int d^2x \left[ \bar{\Psi} \left( i \Gamma^\mu \partial_\mu \right) \Psi + J' \Bar\Psi \tau^a \Psi T^a \delta(x) \right],
    \label{Seff2d}
\end{equation}
where the two-dimensional Dirac matrices are given by $\Gamma^0 = \sigma^1$, $\Gamma^z = -i\sigma^2$ and $J'$ is the effective Kondo coupling.  $\Psi$ is a two-component quark field. In Ref.~\cite{kimura2019conformal}, this model Eq.~\eqref{Seff2d} is studied using the conformal field theory technique. In this approach, the eigenstate of the bulk Hamiltonian with $U(3f)$ symmetry admits a conformal embedding of the form~\cite{parcollet1998overscreened}
\begin{equation}
        SU(3)_f\times SU(f)_3\times U(1)
\end{equation}
and the impurity only couples to the $SU(3)_f$ color degree of freedom, where, using the standard fusion rule, one can explicitly obtain physical quantities related to the impurities like impurity entropy, low-temperature susceptibility, and specific heat~\cite{ludwig1994exact,affleck1991kondo}.

In this work, we investigate the interplay between two competing phenomena—color superconductivity and the color–Kondo effect—within a \(1+1\)D toy model of QCD based on the \(\mathrm{SU}(3)\) Gross–Neveu model~\cite{gross1974dynamical,andrei1980derivation} with an impurity. 
{The Gross–Neveu model was originally introduced as a renormalizable analogue of the NJL model that dynamically generates a fermion mass through an attractive quark–antiquark interaction (chiral channel). 
By performing a Fierz transformation, the same four–fermion interaction can equivalently be expressed in the particle–particle (Cooper) channel, describing pairing correlations between quarks of opposite chirality. 
In this work, we adopt this particle–particle interpretation, so that the dynamically generated mass gap \(m\) is viewed as a superconducting correlation gap arising from antisymmetric color \(\bar{\mathbf{3}}\) pairing.} 
Because the model is \(1+1\)–dimensional, true spontaneous symmetry breaking is forbidden by Coleman’s theorem. 
{Instead, the system exhibits quasi–long–range color–superconducting correlations with algebraically decaying order–parameter correlations, while the mass gap is generated dynamically through dimensional transmutation.} 
When a localized color impurity is placed in the system, a color–Kondo effect takes place as discussed in terms of RG analysis. 
The rest of the paper shows that, while the RG description is correct in the regime \(\Lambda_K \gg \Delta\), the exact solution reveals that the model is significantly richer, exhibiting additional impurity phases.

\section{The Model}
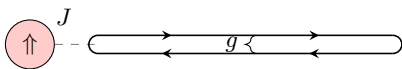
\begin{figure}[H]
    \centering
    \begin{tikzpicture}[scale=1.2, >=stealth]
      \def\n{1}               
      \def\radius{4}           
      \def\gap{0.75}           
      \def\offset{0.1}         
      \def\wirethickness{0.6pt}
      
      \node[draw,circle,fill=red!20,minimum size=6mm] (imp) at (0,0) {$\Uparrow$};
      
      \foreach \i in {1,...,\n} {
        \pgfmathsetmacro{\angle}{360/\n * (\i - 1)}
        \pgfmathsetmacro{\xstart}{\gap * cos(\angle)}
        \pgfmathsetmacro{\ystart}{\gap * sin(\angle)}
        \pgfmathsetmacro{\xend}{\radius * cos(\angle)}
        \pgfmathsetmacro{\yend}{\radius * sin(\angle)}
        \pgfmathsetmacro{\perpx}{-\offset * sin(\angle)}
        \pgfmathsetmacro{\perpy}{\offset * cos(\angle)}
        \coordinate (upStart) at ({\xstart+\perpx}, {\ystart+\perpy});
        \coordinate (upEnd) at ({\xend+\perpx}, {\yend+\perpy});
        \coordinate (lowStart) at ({\xstart-\perpx}, {\ystart-\perpy});
        \coordinate (lowEnd) at ({\xend-\perpx}, {\yend-\perpy});
        
        \draw[line width=\wirethickness,
              decoration={markings,
                  mark=at position 0.25 with {\arrow{>}},
                  mark=at position 0.75 with {\arrow{>}}},
              postaction={decorate}]
              (upStart) -- (upEnd);
        
        \draw[line width=\wirethickness,
              decoration={markings,
                  mark=at position 0.25 with {\arrow{<}},
                  mark=at position 0.75 with {\arrow{<}}},
              postaction={decorate}]
              (lowStart) -- (lowEnd);
        
        \coordinate (gapMid) at ($(imp)!0.5!(\xstart,\ystart)$);
        \pgfmathsetmacro{\labelOffsetX}{-0.3 * sin(\angle)}
        \pgfmathsetmacro{\labelOffsetY}{0.3 * cos(\angle)}
        \node at ($(gapMid) + (\labelOffsetX, \labelOffsetY)$) {\small$J$};
        
        \draw[dashed,gray] (imp) -- (\xstart, \ystart);
        
        \pgfmathsetmacro{\midFrac}{0.5}
        \coordinate (midUp) at ($(upStart)!{\midFrac}!(upEnd)$);
        \coordinate (midLow) at ($(lowStart)!{\midFrac}!(lowEnd)$);
        
        \draw[decorate,decoration={brace,amplitude=4pt,raise=-4pt}]
          (midLow) -- (midUp);
        
        \coordinate (braceMid) at ($(midLow)!0.5!(midUp)$);
        \node[inner sep=1pt] at ($(braceMid)!0.15cm!(0,0)$) {\small$g$};
        
        \pgfmathsetmacro{\dxOuter}{(\xend+\perpx)-(\xend-\perpx)}
        \pgfmathsetmacro{\dyOuter}{(\yend+\perpy)-(\yend-\perpy)}
        \pgfmathsetmacro{\radiusOuter}{sqrt(\dxOuter*\dxOuter + \dyOuter*\dyOuter)/2}
        \pgfmathsetmacro{\angleOuter}{atan2(\dyOuter,\dxOuter)}
        \draw[line width=\wirethickness]
          (upEnd) arc[start angle=\angleOuter, delta angle=-180, radius=\radiusOuter cm];
        
        \pgfmathsetmacro{\dxInner}{(\xstart-\perpx)-(\xstart+\perpx)}
        \pgfmathsetmacro{\dyInner}{(\ystart-\perpy)-(\ystart+\perpy)}
        \pgfmathsetmacro{\radiusInner}{sqrt(\dxInner*\dxInner + \dyInner*\dyInner)/2}
        \pgfmathsetmacro{\angleInner}{atan2(\dyInner,\dxInner)}
        \draw[line width=\wirethickness]
          (lowStart) arc[start angle=\angleInner, delta angle=-180, radius=\radiusInner cm];
        
        \pgfmathsetmacro{\numRad}{\radius + 0.2}
        \pgfmathsetmacro{\nx}{\numRad*cos(\angle) - 0.15*sin(\angle)}
        \pgfmathsetmacro{\ny}{\numRad*sin(\angle) + 0.15*cos(\angle)}
      }
    \end{tikzpicture}   \caption{Cartoon depiction of the model Eq.~\eqref{modelham} where left and right moving chiral fermions with marginal bulk attractive current-current interaction $g$ and open boundary conditions are coupled to an SU(3) moment at the boundary with Kondo coupling $J$.}
    \label{fig:continuum}
\end{figure}

As a first step towards a controlled model, we shall consider a model with a single flavor of quark with attractive interaction between the quarks of opposite chirality, which dynamically opens a mass gap, and we shall consider a single heavy quark impurity localized at $x=0$ where, when the bulk scatters off the impurity, its chirality changes. This is pictorially shown in Fig.~\ref{fig:continuum}. A more realistic description of the QCD Kondo problem in superconducting quark matter by considering a multiflavor problem with attractive color interaction between different intra- and inter-flavors will be presented elsewhere. The Hamiltonian of the model under consideration is
\begin{align}
    H&=-i\int_{-L}^0 \mathrm{d} x \Big\{ \psi^\dagger_{+,a}(x)\partial_x \psi_{+,a}(x)-\psi^\dagger_{-,a}(x)\partial_x \psi_{-,a}(x)   \Big\}\nonumber\\
    &+ \int_{-L}^0 \mathrm{d}x \Big \{g\,\psi^\dagger_{+,a}(x)\vec \tau_{ab} \psi_{+,b} (x)\,\psi^\dagger_{-,c} (x)\vec\tau_{cd}\psi_{-,d}(x)\nonumber\\
    &+ J\psi^\dagger_{+,a}(x)\vec\tau_{a,b}\psi_{-,b}(x)\cdot \vec T \delta(x)\Big\},
    \label{modelham}
\end{align}
where $\psi_{\pm,a}$ is a fermionic annihilator operator where the first index $\pm$ denotes its chirality whereas the second index $a$ is the $SU(3)$ color index. The matrices $\tau$  are Gell-Mann matrices, $SU(3)$   generators that act on the color space of the quarks, and $\vec T$ describes the localized impurity. The coupling constant $g$ is the attractive interaction between the bulk quarks of opposite chirality, whereas $J$ is the Kondo coupling. 

{The bulk interaction 
\[
g\,(\psi_{+}^{\dagger}\tau^{a}\psi_{+})(\psi_{-}^{\dagger}\tau^{a}\psi_{-})
\]
can be Fierz–rearranged into the particle–particle (Cooper) channel, describing pairing correlations between opposite chiralities.\footnote{%
Using the SU(3) Fierz identity 
\((\tau^{a})_{ij}(\tau^{a})_{kl}
= \tfrac{2}{3}(S_{s})_{ik}(S_{s})_{lj}
- \tfrac{4}{3}(A_{a})_{ik}(A_{a})_{lj}\),
one obtains 
\((\psi_{L}^{\dagger}\tau^{a}\psi_{L})(\psi_{R}^{\dagger}\tau^{a}\psi_{R})
= \tfrac{2}{3}(\psi_{L}^{\dagger}S_{s}\psi_{R}^{\dagger})(\psi_{R}S_{s}\psi_{L})
- \tfrac{4}{3}(\psi_{L}^{\dagger}A_{a}\psi_{R}^{\dagger})(\psi_{R}A_{a}\psi_{L})\),
showing that the antisymmetric color \(\bar{\mathbf{3}}\) channel is attractive. 
In \(1+1\)~dimensions, this attraction generates a finite mass gap and algebraically decaying pairing correlations rather than a true condensate.}
In \(\mathrm{SU}(3)\), this decomposition yields symmetric (\(\mathbf{6}\)) and antisymmetric (\(\bar{\mathbf{3}}\)) color combinations, of which the antisymmetric channel is attractive for \(g>0\). 
Although the theory is \(1+1\)-dimensional, where continuous symmetries cannot break spontaneously, the interaction drives quasi-long-range color-superconducting correlations and dynamically generates a mass gap \(m\) through dimensional transmutation.
}

We will describe the various phases of the Hamiltonian as functions of the couplings $J>0$ and $g>0$.
The bulk coupling constant $g$ flows to strong coupling where the model opens a superconducting gap (interpreted here as the dynamical mass generated by the Gross–Neveu mechanism in the particle–particle channel)~\cite{gross1974dynamical,andrei1980derivation}. In the absence of $g$, the Kondo coupling $J$ also flows to strong coupling where the impurity is screened by a multiparticle cloud of bulk fermions~\cite{jerez1998solution,andrei1983solution}. In this work, we shall provide an exact solution for the Hamiltonian, which allows the study of the interplay between the two phenomena when both couplings are present. 

This interplay between superconductivity and the Kondo screening has long been a subject of study in condensed matter physics~\cite{Yu,Shiba,Rusinov,sakurai1970comments,franke2011competition,PhysRevLett.121.207701,aoi1974magnetic,PhysRevLett.26.428,matsuura1977effects,takano1969kondo,cuevas2001kondo}. In a superconducting material, the fundamental state is characterized by the attractive interaction between electrons, leading to the formation of Cooper pairs and a globally coherent superconducting phase~\cite{bardeen1957microscopic}. However, the introduction of a magnetic impurity can alter this picture locally. Instead of participating in the collective Cooper pairing, conduction electrons in the vicinity of the impurity become strongly correlated with it, forming a localized many-body singlet state analogous to the Kondo effect observed in normal metals containing magnetic impurities.  

The competition between the bulk's tendency to form Cooper pairs and the local tendency to screen the impurity through singlet formation gives rise to a fascinating array of phenomena in the vicinity of the impurity~\cite{kattel2024overscreened,pasnoori2020kondo,PhysRevB.105.174517}. Depending on the relative strength of the coupling between the bulk superconducting electrons $g$ and the impurity $J$, the impurity's magnetic moment can be screened by a complex multi-particle cloud, characteristic of the Kondo effect. Or the impurity can induce localized bound states within the superconducting energy gap, known as Yu-Shiba-Rusinov (YSR) states, which effectively screen the impurity through single-particle excitations~\cite{Yu,Shiba,Rusinov}. In yet other regimes, the impurity might remain entirely unscreened, retaining its local magnetic moment.

Equation~\eqref{modelham} captures, within quark matter, a competition akin to the interplay between the Kondo effect and superconductivity in condensed-matter physics. Concretely, massless quarks at the ultraviolet scale undergo nonperturbative dynamics in the infrared, leading both to gap formation and to the emergence of a superconducting bulk state. Introducing a heavy quark impurity at the edge of this bulk then acts as a local perturbation. This perturbation can trigger intriguing boundary physics, as the impurity interacts with the surrounding superconducting quarks. Much like the competition between Cooper pairing and Kondo singlet formation in conventional superconductors, the interplay between color superconductivity in the quark bulk and the QCD Kondo effect associated with the impurity leads to a rich phase diagram. Within this framework, the heavy quark impurity can be screened by bulk fermionic excitations or through the formation of localized subgap states, analogous to the Yu-Shiba-Rusinov states. The ultimate fate of the impurity – whether it becomes decoupled, or is screened by the bulk, or forms a bound state within the superconducting gap – is dictated by the relative strengths of the color superconducting and  Kondo interactions. This intricate competition not only enriches our understanding of the QCD phase diagram but also provides valuable insights into impurity physics under extreme conditions, potentially relevant to phenomena observed in high-energy heavy-ion collisions and within the dense interiors of compact stars. The adaptation of theoretical methods developed in condensed matter physics offers a promising avenue for further exploration of these exotic states of matter.

Before presenting our main results, we note that recent advances in cold atom technology make it plausible to realize the toy model proposed here in the laboratory.  In particular, the ability to engineer and probe local moments in multicomponent fermionic systems—thereby accessing a Kondo‐like coupling~\cite{amaricci2025engineering}—and the development of SU($N$)‐symmetric gases using alkaline-earth-like atoms have been demonstrated in Refs.~\cite{Taie2010,Scazza2014,Pagano2014}.  Thus, it should be possible to implement our model within a suitably designed cold atom experiment in the near future by realizing the setup shown in schematic Fig.\ref{fig:continuum}.

\section{Summary of main results}
In the language of the renormalization group (RG), the model described in Eq.~\eqref{modelham} exhibits an interplay between bulk and boundary interactions, characterized by two coupling constants $g$ and $J$, both of which flow under the action of RG. The bulk $\beta$-function, given by
\begin{equation}
    \beta(g) = \frac{dg}{d\ln D} = -\frac{12}{\pi} g^2,
    \label{ggflow}
\end{equation}
indicates that the bulk coupling $g$ grows as the energy cut-off $D$ is lowered. This infrared divergence and UV asymptotic freedom signify that the model flows to a strong coupling, which is characterized by the non-perturbative generation of a mass gap within the SU(3) color sector of the theory. In other words, the Hamiltonian Eq.~\eqref{modelham} contains no bare mass scales. In the ultraviolet its color sector is described by the massless $SU(3)_1$ Wess–Zumino–Witten CFT. This conformal point is perturbed by the marginally relevant operator $J_L^aJ_R^a$, whose one‐loop $\beta$‐function $\beta(g)\propto - g^2$
drives the coupling to the strong coupling in the infrared. To quantize such a renormalizable theory, we introduce an ultraviolet cutoff $D$ and let the bare coupling $g(D)$ run so that the combination
$m=D e^{-1/g(D)}$
remains finite as $D\to\infty$. By dimensional transmutation, this fixes the mass scale $m$, gapping out all nonabelian (color) excitations and dynamically generating the superconducting gap~\cite{gross1974dynamical}.

  At the boundary $x=0$, the impurity $\vec T$ is coupled to the bulk through the color-Kondo term, which alone would flow to strong coupling and be fully screened in the absence of bulk superconducting order~\cite{andrei1983solution}. More concretely, in the absence of bulk coupling $g$, the scaling dimension of the Kondo perturbation is $\Delta=1$, and hence the beta function vanishes at tree level and one finds $\beta(J)=-CJ^2$ in one loop with $C>0$ being a positive constant. Since the beta function is negative, the Kondo coupling flows to a strong coupling limit. However, as we shall see from the exact solution, in the presence of the bulk coupling $g$, the Kondo interaction picks up the anomalous dimension of the form $1+a g$ where $a$ is a positive constant, and hence the RG flow of the boundary operator becomes $\beta(J)=-C J^2+ a g J$ which shows that the contribution from the bulk coupling $g$ to the boundary flow is positive. In other words, this contribution tries to decrease the strength of the Kondo coupling and hence tends to unscreen the impurity. These two competing terms in the presence of the bulk order give rise to three distinct IR boundary phases: (i) an $SU(3)$-symmetric Kondo phase in which the impurity is screened by a many-body cloud of quasiparticles; (ii) a Yu–Shiba–Rusinov–like phase where a single bound quasiparticle forms a screening singlet with the quark; and (iii) an unscreened local-moment phase in which the impurity remains effectively unscreened. We determine the constants $C$ and $a$ from the exact non-perturbative  $\beta$-function $\beta(J) = \frac{dJ}{d\ln D}$ obtained from the Bethe Ansatz methods in our cut-off scheme, which takes the form
{\begin{equation}
   \beta(J)\approx -\frac{6 J}{\pi } (J-2 g)
   \label{jjflow}
\end{equation}}
at an initial ultraviolet (UV) energy scale $D_0$. As discussed above, it indeed comprises two competing effects: a standard Kondo-type term, $-CJ^2$ with $C=\frac{6}{\pi}$, which tends to drive $J$ towards the strong coupling, and a bulk-induced term, $+a g J$ with $a=\frac{12}{\pi}$, which acts to suppress $J$. The fact that the bulk coupling $g$ flows to strong coupling as the energy scales decrease, coupled with the boundary coupling $J$ exhibiting signs and magnitudes that are intricately tied to both $g$ and its own value, unveils a far more nuanced boundary physics than the standard Kondo effect in Fermi liquid theory. This interplay of bulk and boundary RG flow shapes the fate of impurity, leading to a richer tapestry of potential behaviors. 

The condition $J = 4g$ on this initial scale defines a critical line that acts as a separatrix that delineates two distinct pathways for the boundary interaction as the energy scale is reduced. If the initial boundary coupling $J_0$ starts above this line, such that $J_0 > 4g_0$, the beta function $\beta(J)$ is positive, indicating that $J$ will flow towards stronger coupling as the bulk mass gap develops. This strong coupling regime ultimately leads to the complete Kondo screening of the impurity by the bulk fermions, resulting in the formation of an SU(3) singlet state. In contrast, if the initial boundary coupling $J_0$ lies below the separatrix, with $J_0 < 4g_0$, the beta function $\beta(J)$ becomes negative. This negative flow drives the boundary coupling $J$ toward zero as the energy scale is lowered. Consequently, when the bulk mass gap eventually opens up in the infrared, the impurity remains essentially decoupled from the bulk and unscreened. Thus, this weak coupling renormalization group (RG) analysis reveals a crucial condition for impurity screening: the initial exchange coupling $J_0$ must exceed quadruple the initial bulk coupling $g_0$.

While this analysis sheds light on the initial flow of the couplings, it is important to recognize that it is based on a weak-coupling approximation and does not provide a complete picture of the system's behavior across all energy scales. To fully understand the infrared (IR) theory and the ultimate fate of the impurity, a more comprehensive strong-coupling analysis is required. To achieve this, we shall employ the Bethe Ansatz method, which allows for an exact solution of the model in the presence of both bulk and boundary interactions. This approach will provide a deeper understanding of the various phases and the low-energy behavior of the system.

\begin{figure}[H]
    \centering    \includegraphics[width=0.9\linewidth]{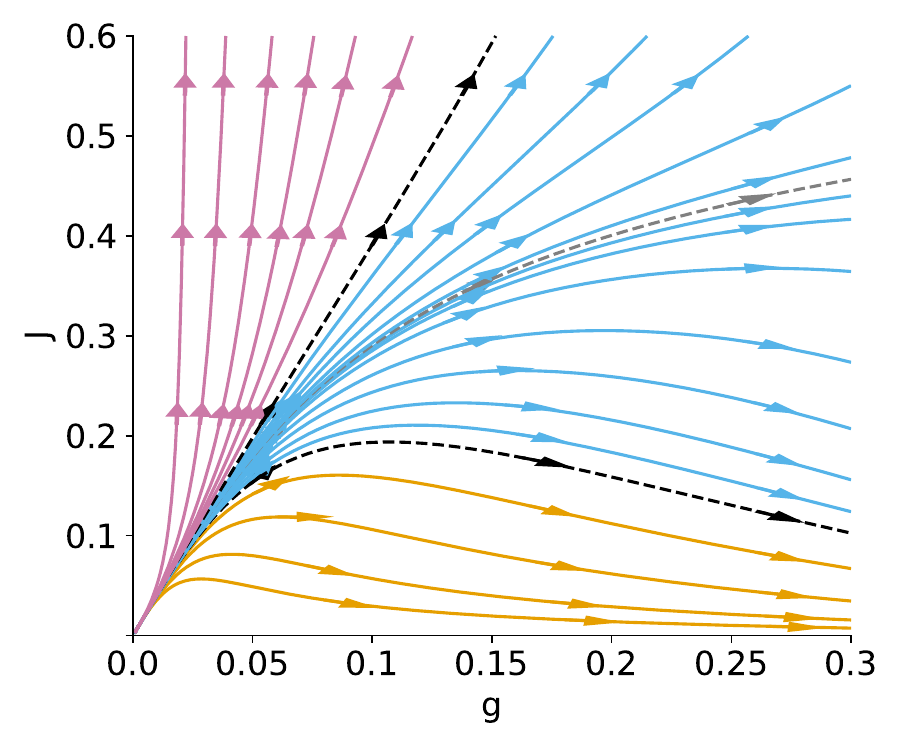} 
    \caption{The weak-coupling RG flow diagram for Hamiltonian Eq.~\eqref{modelham} is given by two RG equations Eq.~\eqref{ggflow} and Eq.~\eqref{jjflow}. The flow is away from the non-interacting point $J=0=g$. The magenta curves denote the Kondo phase, where the boundary coupling flows to strong coupling, and the impurity is screened by a multiparticle Kondo cloud. The cyan lines indicate the YSR phase, where the impurity is screened by a single-particle bound mode. Finally, the orange lines depict the unscreened regime, where the boundary coupling flows to zero, leaving the impurity unscreened. The dashed black curves demarcate the three boundary phases of the model. {The dashed gray curve within the cyan YSR region indicates the locus where the energy of the boundary bound state $E_{\delta}$ vanishes, corresponding to the level crossing at $\delta = 5/4$ shown in Fig.~\ref{fig:Eng-bm-delta}. Across this line, the bound-state energy changes sign, separating the 
screened ($E_{\delta}<0$) and unscreened ($E_{\delta}>0$) regimes within the YSR phase.}}
    \label{fig:RGplot-plot}
\end{figure}

The full renormalization group picture, as interpreted from the exact solution obtained via the Bethe Ansatz, is visually summarized in the RG flow diagram presented below (Fig.~\ref{fig:RGplot-plot}). This diagram illustrates the flow of the couplings and demarcates the distinct regions corresponding to different impurity behaviors. As anticipated from the weak-coupling analysis, we observe regimes where for $J_0 \gg g_0$, the impurity is indeed screened by a multi-particle Kondo cloud in the infrared, corresponding to a flow towards strong coupling. Similarly, for $J_0 \ll g_0$, the impurity remains a local moment, consistent with the boundary coupling flowing toward zero. Crucially, the exact solution unveils an additional region in the RG flow, emerging when $J_0 \approx g_0$. In this intermediate regime, the impurity is screened by a single-particle bound mode that forms below the bulk mass gap, akin to the Yu-Shiba-Rusinov states observed in conventional superconductors. This rich structure in the RG flow, encompassing regions indicative of Kondo screening, an unscreened local moment, and single-particle screening, underscores the complex interplay between the dynamically generated bulk properties and the boundary impurity, providing a more complete understanding than the initial weak-coupling RG analysis could offer.

The renormalization group flow diagram depicted in Fig.~\ref{fig:RGplot-plot} illustrates the evolution of the couplings $g$ and $J$ as the energy scale is reduced, originating from the non-interacting fixed point at $J = g = 0$. These RG equations, derived using an unconventional cutoff scheme within the framework of the Bethe Ansatz solution and detailed in the Appendix~\ref{RGeqn-deriv}, reveal distinct regions corresponding to different impurity behaviors. The flow lines shown in {orange} represent the parameter regimes where the boundary coupling $J$ flows toward weaker values in the infrared, ultimately leading to the impurity remaining as a local, unscreened moment. In contrast, the flow lines in magenta correspond to the parametric regimes where the boundary coupling $J$ flows towards strong coupling, resulting in the screening of the impurity by a Kondo cloud of multiple particles. In particular, the flow lines in cyan illustrate the parametric regimes where the boundary coupling evolves towards the YSR phase characterized by screening the impurity through the formation of a single particle bound mode below the bulk mass gap.

\section{Summary of Phases}

In the previous section, we discussed the existence of the various boundary phases in terms of the two couplings, the bulk superconducting coupling $g$ and the boundary Kondo coupling $J$ both flowing under the action of RG. However, to analyze the phases more concretely, we introduce the RG-invariant combination
\begin{equation}
    d^2 = b^2 - \frac{2b}{c} - 1,
\end{equation}
where $b = \frac{9 - 32g^2}{72g} $ denotes the running bulk coupling and $ c = \frac{36J}{9 - 32J^2} $ is the corresponding running boundary coupling. This RG invariant parameter $d$ will serve as our primary parameter for characterizing the different phases. Since $\{g,J\}\ge 0$, we have $d^2\in\mathbb{R}$, so $d$ is either real or purely imaginary. In the latter case, we introduce a parameter $\delta$ by
$
d = i\delta,\quad \delta\in\mathbb{R}
$. As previously noted, the precise dependence of the coupling constants $g$ and $J$ on the boundary rapidity shift $d$ emerges from the non-perturbative Bethe Ansatz solution, which introduces a non-trivial cutoff directly on the interacting (dressed) momenta. This approach stands in contrast to conventional quantum field theory methods, where regularization is typically imposed on the free (quadratic) part of the action \footnote{We also stress that the running coupling constant in the Bethe Ansatz formalism is non-analytically related to the running coupling constant in standard UV cutoffs. While this implies that the renormalization group (RG) flows derived in the two frameworks are parametrized differently, we emphasize that the beta function computed from the Bethe Ansatz matches the perturbative QFT result in the limit of small bare coupling, thereby confirming consistency between the two approaches in the weak-coupling regime. (See Ref.~\cite{andrei1983solution} for a detailed discussion of the two cut-off schemes.)}.
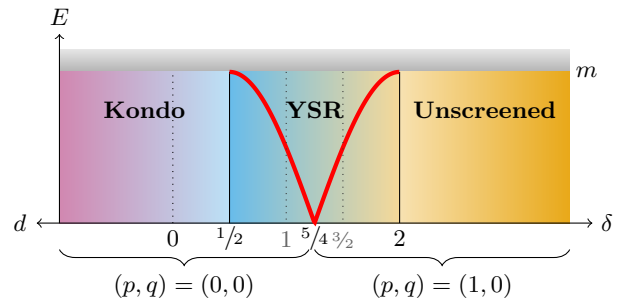
\begin{figure}
    \centering
\begin{tikzpicture}[xscale=1.5]
  \usetikzlibrary{decorations.pathreplacing}

  \definecolor{phaseK}{HTML}{CC79A7}
  \definecolor{phaseY}{HTML}{56B4E9}
  \definecolor{phaseU}{HTML}{E69F00}
  \definecolor{cont1}{gray}{0.9}
  \definecolor{cont2}{gray}{0.7}

  \shade[top color=cont1, bottom color=cont2] (-1,2) rectangle (3.5,2.3);

  \shade[left color=phaseK!90, right color=phaseY!40] (-1,0) rectangle (0.5,2);
  \shade[left color=phaseY!90, right color=phaseU!40] (0.5,0) rectangle (2,2);
  \shade[left color=phaseU!30, right color=phaseU!90] (2,0) rectangle (3.5,2);

  \draw (0.5,0) -- (0.5,2);
  \draw (2,0)   -- (2,2) ;
  \draw[dotted] (0,0) -- (0,2);
  \node at (0,-0.2) {$0$};
  \node at (0.5,-0.2) {$\sfrac{1}{2}$};
  \node at (1.25,-0.2) {$\sfrac{5}{4}$};
  \node at (2,-0.2) {$2$};

  \draw[<->] (-1.2,0) -- (3.7,0) node[right] {$\delta$};
  \draw[->]  (-1,0)   -- (-1,2.5) node[above] {$E$};
  \node at (-1.35,0) {$d$};
  \node at (3.65,2) {$m$};

  \node[font=\small\bfseries] at (-0.25,1.5) {Kondo};
  \node[font=\small\bfseries] at (1.25,1.5) {YSR};
  \node[font=\small\bfseries] at (2.75,1.5) {Unscreened};

  \draw[red,ultra thick]
    plot[domain=0.5:2, samples=200]
      (\x, {2 * abs(sin(30*(4*\x + 1)))});

  \draw[decorate, decoration={brace, mirror, amplitude=6pt}]
    (-1,-0.35) -- (1.2,-0.35)
    node[midway, below=6pt] {$(p,q)=(0,0)$};
  \draw[decorate, decoration={brace, mirror, amplitude=6pt}]
    (1.25,-0.35) -- (3.5,-0.35)
    node[midway, below=6pt] {$(p,q)=(1,0)$};
      \draw[dotted, black!80] (1,2)   -- (1,0) node[below, black!60] {\small $1$};
      \draw[dotted, black!80] (1.5,2)   -- (1.5,0) node[below, black!60] {\small $$\sfrac{3}{2}$$};
\end{tikzpicture}
    \caption{Schematic phase diagram of the SU(3)–Gross–Neveu model with a massive quark impurity at one edge. The low-lying spectrum is plotted as a function of the RG-invariant $d$, which, when imaginary, is written $d = i\delta$. The YSR state exists only for $\frac12 < \delta < 2$, and the magnitude of its energy is shown by the solid magenta line. A first-order quantum phase transition between the screened $(p,q)=(0,0)$ and unscreened $(p,q)=(1,0)$ impurity configurations occurs at $\delta = \frac54$. For $\delta > 2$, the impurity remains unscreened, and no YSR states screening the impurity are present. In the Kondo regime, $\delta < \frac12$ (i.e. $d\in\mathbb{R}$), many-body screening generates a Kondo scale $T_K$. The shaded orange region denotes the continuum of bulk excitations above the mass gap $m$. The two faint vertical guides at $\delta=1$ and $\delta=\tfrac{3}{2}$ delimit the parametric range in which a midgap antiquark excitation carrying triality 2 exists with its energy in range $\frac{m}{2}<E_\delta<m$, whereas for $\delta>\frac{3}{2}$ its energy is $E(\theta)=m\cosh\left(\frac{2\pi}{3}\theta\right)$ such that it merges into the continuum of excitations $E>m$.}
    \label{fig:phase-diagram}
\end{figure}

We proceed to discuss the various phases the system possesses, as functions of the RG invariant parameter $d$. As shown in Fig.~\ref{fig:phase-diagram}, for all values of $d\in \mathbb{R}$ and for $0<\delta<\frac{1}{2}$   the impurity quark is completely screened by a multiparticle cloud of bulk superconducting quarks. In this regime, the Kondo fluctuations overwhelm the bulk superconducting order, and in addition to the bulk mass gap $\Delta$, a new scale $T_K$, the Kondo temperature,  is dynamically generated, which characterizes the screening of the impurity quark and the local impurity density of states, which attains the familiar Lorentzian form. As $\delta$ increases past $\frac12$, superconducting pairing increases and competes with Kondo screening, causing a bound state to emerge below the bulk mass gap. In the range $\frac12<\delta<\frac54$, the bound‐state energy is negative, and the impurity is screened by a single‐particle mode. At $\delta=\frac54$, the bound level crosses zero, and for $\frac54<\delta<2$, its positive energy makes the unscreened impurity lower in energy, marking a first-order quantum phase transition. Beyond $\delta>2$, superconductivity wins over the Kondo scale, and the impurity remains effectively unscreened.

In the next sections, we will use the Bethe Ansatz to study the low-energy physics of our model in each of its three boundary phases. We will show that all color excitations are gapped, while the $U(1)$ charge excitations remain gapless. As we vary the RG-invariant parameter $d$, the system moves between different boundary phases. We will examine the low-energy excitations in each phase. As shown in Ref.\cite{andrei1980derivation}, the Hamiltonian \eqref{modelham} is integrable in the $J\to0$ limit, and Ref.\cite{andrei1983solution} demonstrates integrability in the $g\to0$ limit. In Appendix~\ref{npart-state} we construct the generic $N_q$‐particle eigenstates. We then show in Appendix~\ref{sec:int} that the model remains integrable when both $g$ and $J$ are nonzero, and in Appendix~\ref{sec:BAE-deriv} we derive the Bethe Ansatz equations. Moreover, the Bethe Ansatz equations are explicitly solved in Appendix~\ref{sec:BAE-soln}. Here, we shall summarize the results in all the phases in great detail.

\section{Bethe Ansatz Equations}

{The Bethe Ansatz~\cite{bethe1931theorie,yang1967some,gaudin1967systeme,bergknoff1979structure,andrei1980derivation} provides an exact and algebraic procedure for diagonalizing a class of integrable interacting Hamiltonians. 
Its central idea is that the many-body wavefunction can be expressed as a coherent superposition of plane waves, with amplitudes related by two-body scattering matrices that satisfy the Yang--Baxter equation. 
These two-body \(S\)-matrices ensure that all multiparticle scattering processes factorize into products of two-body collisions, making the problem exactly solvable. 
Imposing either periodic or open boundary conditions quantizes the particle rapidities (or quasimomenta), leading to a discrete set of algebraic equations---the Bethe Ansatz equations. 
In models with internal symmetry, the number of rapidity families equals the rank of the underlying Lie algebra.}

After introducing the ultraviolet cutoff \(D\) and finite volume \(L\) to convert the field-theoretic eigenvalue problem into an \(N\)-body Hamiltonian, 
we find that diagonalizing the Hamiltonian reduces to determining discrete quasimomenta \(\{k_j\}\) that satisfy two coupled algebraic Bethe equations. 
The two distinct sets of rapidities, \(\{\lambda_\alpha\}\) and \(\{\mu_k\}\), naturally arise from the rank-2 structure of the \(\mathrm{SU}(3)\) algebra. 
This directly corresponds to the two simple roots of the \(A_2\) Dynkin diagram, dictating the form of the Bethe Ansatz equations. 
{Solving these coupled equations yields the complete energy spectrum and identifies each multiplet with an irreducible representation of \(\mathrm{SU}(3)\); further technical details of the nested construction are provided in Appendices~\ref{npart-state}, \ref{sec:int} and \ref{sec:BAE-deriv}.}

We pause here to briefly summarize some relevant facts of the representation theory of the $SU(3)$ algebra, which are reflected in the Bethe Ansatz solution. In SU(3), every irreducible representation is uniquely specified by two non-negative integers $(p,q)$, its Dynkin labels, which determine both its dimension
\[
\dim(p,q)=\frac{(p+1)(q+1)(p+q+2)}{2}
\]
and its highest-weight vector $\Lambda=p\lambda_1+q\lambda_2$.  The highest-weight state then has the third isospin
$
I_3^{\max}=\frac{p}{2},
$
and hypercharge
$
Y^{\max}=\frac{p+2q}{3},
$
and all other weights $(I_3,Y)$ in the multiplet, followed by subtracting integer combinations of the simple roots.  Since the center of SU(3) is the cyclic group $\{1,\omega,\omega^2\}$ with $\omega=e^{2\pi i/3}$, each $(p,q)$ irrep also carries a discrete $\mathbb{Z}_3$ triality
\[
t\equiv p+2q\pmod3,
\]
which labels its phase $\omega^t$ under the central element.  We will, therefore, label any SU(3) state by its Dynkin labels $(p,q)$, by its weight $(I_3,Y)$ within that multiplet, and by the discrete $\mathbb{Z}_3$ triality $t$.

The quantization conditions for the discrete quasimomenta for the theory defined on a line segment $x\in [-L,0]$ are given by
\begin{equation}
\label{BAE-quasi-momenta}
e^{-2i k_j L} = \prod_{\alpha=1}^{M_1} \prod_{\upsilon=\pm}
\left( \frac{b + \upsilon \lambda_\alpha + \frac{i}{2}}{b + \upsilon \lambda_\alpha - \frac{i}{2}} \right)
\end{equation}
where the roots $\lambda_\alpha$ together with the roots $\mu_k$ satisfy the rank-1 Bethe Ansatz equations of the form
\begin{equation}
\label{bae:rank1-eqn}
\begin{aligned}
&\prod_{\upsilon=\pm} \left[
\left( \frac{\lambda_\alpha + \upsilon b - \frac{i}{2}}{\lambda_\alpha + \upsilon b + \frac{i}{2}} \right)^{N_q}
\frac{\lambda_\alpha + \upsilon d - \frac{i}{2}}{\lambda_\alpha + \upsilon d + \frac{i}{2}}
\prod_{k=1}^{M_2}
\frac{\lambda_\alpha + \upsilon \mu_k - \frac{i}{2}}{\lambda_\alpha + \upsilon \mu_k + \frac{i}{2}}
\right]\\
&=
\quad\quad\prod_{\substack{\beta=1 \\ \beta \ne \alpha}}^{M_1}
\prod_{\upsilon=\pm}
\frac{\lambda_\alpha + \upsilon \lambda_\beta - i}{\lambda_\alpha + \upsilon \lambda_\beta + i};
\quad \alpha = 1,\dots, M_1
\end{aligned}
\end{equation}
and the rank-2 Bethe Ansatz equations of the form
\begin{equation}
\label{bae:rank2-eqn}
\begin{aligned}
\prod_{\alpha=1}^{M_1} \prod_{\upsilon=\pm}
\frac{\mu_k + \upsilon \lambda_\alpha - \frac{i}{2}}{\mu_k + \upsilon \lambda_\alpha + \frac{i}{2}}
&=
\prod_{\substack{\ell=1 \\ \ell \ne k}}^{M_2}
\prod_{\upsilon=\pm}
\frac{\mu_k + \upsilon \mu_\ell - i}{\mu_k + \upsilon \mu_\ell + i},
\end{aligned}
\end{equation}
where $k=1,\cdots,M_2$.
The two quantum numbers $M_1$ and $M_2$ count the number of first‐ and second‐level magnons in the Bethe Ansatz for the system with $N_q+1$ particles.  One then defines the color occupation numbers
\[
r_1 = (N_q+1) - M_1,\quad
r_2 = M_1 - M_2,\quad
r_3 = M_2,
\]
which satisfy $r_1 + r_2 + r_3 = N_q+1$.  The corresponding $\mathrm{SU}(3)$ ($A_{2}$) Dynkin labels are
\[
p = r_{1} - r_{2} = (N_{q} + 1) - 2M_{1} + M_{2}, 
\quad
q = r_{2} - r_{3} = M_{1} - 2M_{2},
\]
and the pair $(p,q)$ uniquely specifies the representation.

\smallskip

We now turn to the solutions of these equations in the three boundary regimes, the Kondo phase, the YSR phase, and the unscreened phase, presented in order of decreasing detail. 

\subsection{The Kondo Phase}
When the RG invariant parameter $d$ is real, or it takes the purely imaginary values $d=i\delta$ such that $0<\delta<\frac{1}{2}$, the impurity is screened by a multiparticle cloud of massive bulk quarks. In this regime, the scale of boundary interaction is much stronger than the bulk superconducting order and hence the impurity phase is very similar to the massless case where $g\to 0$ which has been intensely studied in literature~\cite{andrei1983solution,jerez1998solution,parcollet1998overscreened,zinn1998generalized,ludwig1994exact}. The impurity quark, which carries an ultraviolet entropy of $\ln 3$, becomes completely screened in the infrared, leading to a vanishing impurity entropy. Here, we study this screening behavior in the presence of a finite but small bulk coupling $g$, compared to the boundary interaction strength $J$. 

To study the IR limit of the model in the Kondo phase, we solve the Bethe Ansatz equations Eq.~\eqref{BAE-quasi-momenta}, Eq.~\eqref{bae:rank1-eqn}, and Eq.~\eqref{bae:rank2-eqn}. Taking the logarithm on both sides of the Bethe Ansatz equation \eqref{BAE-quasi-momenta} for $d\in \mathbb{R}$, we write
\begin{equation}
    k_j=\frac{\pi n_j}{L} + \frac{1}{L} \sum_{\alpha=1}^{M_1} \left[ \tan^{-1}(2(b + \lambda_\alpha)) + \tan^{-1}(2(b - \lambda_\alpha)) \right]
\end{equation}
such that summing over $j$ we obtain the energy
\begin{equation}
    E = \sum_{j} \frac{\pi n_j}{L} + D\sum_{\upsilon=\pm} \sum_{\alpha=1}^{M_1} \left[ \tan^{-1}(2(b +\upsilon \lambda_\alpha))  \right],
    \label{engeqn}
\end{equation}
where $n_j$ are the integers, which are charge quantum numbers. 

Moreover, the logarithm of the two ranks of the Bethe Ansatz equations \eqref{bae:rank1-eqn} and \eqref{bae:rank2-eqn} reads
    \begin{align}
    &\sum_{\upsilon=\pm}\Bigg(N_q \tan^{-1}(2 (\lambda_\alpha+ \upsilon b))+\tan^{-1}(2 (\lambda_\alpha+ \upsilon d
   )) \nonumber\\
   &\left.+\sum_{k=1}^{M_2}\tan^{-1}(2(\lambda_\alpha+\upsilon\mu_k))\right)+\tan^{-1}(2\lambda_\alpha)\nonumber\\
   &\quad\quad\quad=\pi I_\alpha +\sum_{\upsilon=\pm}\sum_{\beta}\tan^{-1}(\lambda_\alpha+\upsilon\lambda_\beta)
   \label{rank1EQN}
\end{align}
and
\begin{align}
    &\tan^{-1}(2\mu_k)+\sum_{\upsilon=\pm} \sum_{\alpha=1}^{M_1}\tan^{-1}(2(\mu_k+\upsilon \lambda_\alpha))\nonumber\\
    &=\sum_{\upsilon=\pm}\sum_{\ell=1}^{M_2}\tan^{-1}(\mu_k+\upsilon\mu_{\ell})+ \pi \mathcal{I}_{k},
    \label{rank2EQN}
\end{align}
where $I_\alpha$ and $\mathcal{I}_k$ are integers that represent the color quantum numbers. To analyze the coupled equations Eq.~\eqref{rank1EQN} and Eq.~\eqref{rank2EQN} in the thermodynamic limit, we introduce the densities of roots as follows

\[
\rho_1(\lambda_j)=\frac{1}{\lambda_{j+1}-\lambda_j},
\qquad
\rho_2(\mu_k)=\frac{1}{\mu_{k+1}-\mu_k}.
\]

such that we can convert all the sums over roots to integrals using the following identities
\begin{align}
   & \sum_{\beta=1}^{M_1}F(\lambda_\beta)\to\int_{-\infty}^\infty F(\lambda')\rho_1(\lambda')d\lambda',\nonumber\\
&\sum_{k=1}^{M_2}G(\mu_k)\to\int_{-\infty}^\infty G(\mu')\rho_2(\mu')d\mu'.\nonumber
\end{align}
Differentiating with respect to $\lambda_\alpha$ (and similarly $\mu_k$) turns each
$\tan^{-1}(x)$ into the Lorentzian $\frac{d}{dx}\tan^{-1}(x)=\frac{1}{1+x^2}$,
and the two equations become
\begin{align}
2\rho_1(\lambda)
&= f(\lambda)
   + \sum_{\upsilon=\pm}
     \int K^{\{2\}}(\lambda+\upsilon\mu)\rho_2(\mu)\mathrm{d}\mu
\notag\\
&\quad
   - \sum_{\upsilon=\pm}
     \int K^{\{1\}}(\lambda+\upsilon\lambda')\rho_1(\lambda')\mathrm{d}\lambda'
   - \delta(\lambda),
\\
2\rho_2(\mu)
&= a_{\frac12}(\mu)
   + \sum_{\upsilon=\pm}
     \int K^{\{2\}}(\mu+\upsilon\lambda)\rho_1(\lambda)\mathrm{d}\lambda
\notag\\
&\quad
   - \sum_{\upsilon=\pm}
     \int K^{\{1\}}(\mu+\upsilon\mu')\rho_2(\mu')\mathrm{d}\mu'
   - \delta(\mu).
\end{align}

The $\delta$-functions in each equation eliminate the spurious roots $\lambda=0$ and $\mu=0$, preventing non-normalizable solutions. Moreover, here
\begin{align}
&f(\lambda)=a_{\frac{1}{2}}(\lambda)+\sum_{\upsilon=\pm}\left(N_q a_{\frac{1}{2}}(\lambda+\upsilon b) +a_{\frac{1}{2}}(\lambda+\upsilon d)\right)\nonumber\\
&a_\gamma(\lambda)= \frac{1}{\pi}\frac{\gamma}{\lambda^2+\gamma^2}\nonumber\\
&K^{\{n\}}(\rho)=a_{\frac{1}{n}}(\rho)\nonumber
\end{align}

Now, we can solve these two equations in Fourier space to obtain the solution of the form
\begin{align}
\tilde{\rho}_1(\omega)&=\frac{\left(2 \left(e^{| \omega | }+1\right) \left(N_q \cos (b \omega )+\cos (d \omega )\right)-e^{\frac{3 |
   \omega | }{2}}+1\right)}{2e^{\frac{| \omega | }{2}}  (2 \cosh ( \omega )+1)}\label{rank1fs}
\\
   \tilde{\rho}_2(\omega)&=\frac{2 \left(N_q \cos (b \omega )+\cos (d \omega
   )\right)+e^{-\frac{| \omega | }{2}}-e^{| \omega | }}{4 \cosh
   (\omega  )+2}\label{rank2fs}
\end{align}

In the ground state, one finds that the first‐rank roots occupy two‐thirds of the total,
\begin{equation}
    M_1 =\int \rho_1(\lambda)d\lambda =\tilde\rho_1(0)
=\frac{2}{3}(N_q+1),
\end{equation}
while the second‐rank roots fill the remaining one‐third,
\begin{equation}
    M_2 =\int \rho_2(\mu)d\mu =\tilde\rho_2(0)
=\frac{1}{3}(N_q+1).
\end{equation}

For the ground state, one obtains
\[
p = (N_q+1) - 2M_1 + M_2 = 0,\qquad
q = M_1 - 2M_2 = 0,
\]
so the representation has Dynkin labels $(p,q)=(0,0)$, i.e.\ the $SU(3)$ singlet. Its dimension is 1
and its $\mathbb{Z}_3$ triality is $t \equiv (p + 2q)\bmod 3 = 0$. Since the ground state is a color singlet, the color charge of the impurity is fully screened.

Notice that when $d$ becomes purely imaginary, we set $d = i\delta$. For $\delta < 1/2$, all of the results above continue to hold by analytic continuation. However, once $\delta > 1/2$, the sign of the imaginary part in the rank-1 Bethe Ansatz equation’s impurity term flips, so the contour deformations and resulting root densities change, and one must redo the analysis to account for that sign reversal.
Thus, the Kondo phase in the model exists when $d$ is real and when $\delta < 1/2$, where the impurity quark is screened by a multiparticle cloud of bulk quarks.

\subsubsection{Fundamental bulk excitations in the Kondo phase}\label{bulk-excitation}

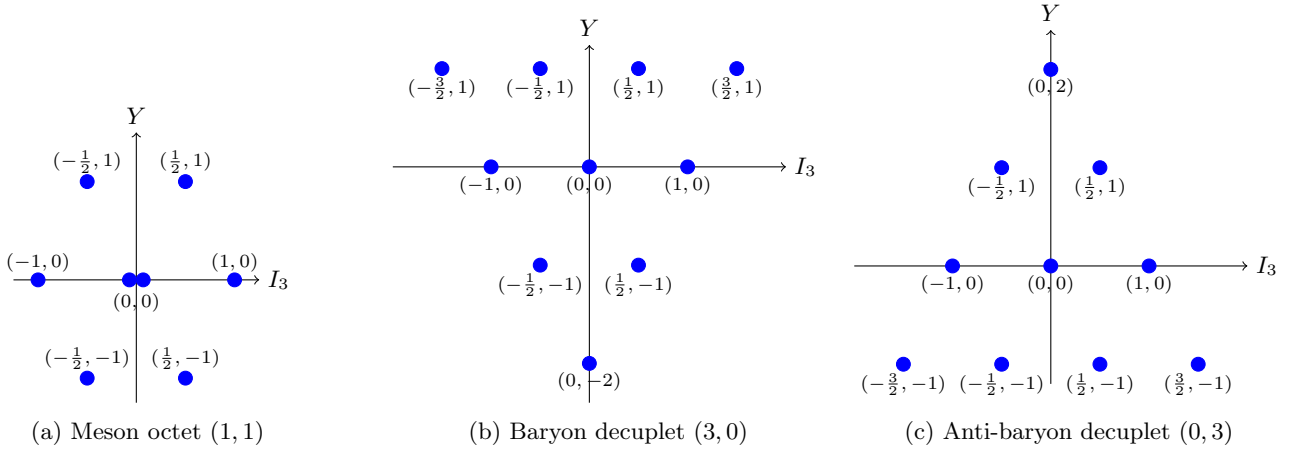
\begin{figure*}[htbp]
    \centering

    \begin{subfigure}[b]{0.32\textwidth}
        \centering
        \begin{tikzpicture}[scale=1.3]
          \draw[->] (-1.25,0) -- (1.25,0) node[right] {$I_3$};
          \draw[->] (0,-1.25) -- (0,1.5) node[above] {$Y$};
          \foreach \x/\y/\label in {
            1/0/{$(1,0)$}, -1/0/{$(-1,0)$},
            0.5/1/{$(\frac12,1)$}, -0.5/1/{$(-\frac12,1)$},
            0.5/-1/{$(\frac12,-1)$}, -0.5/-1/{$(-\frac12,-1)$}}
          {
            \filldraw[blue] (\x,\y) circle (2pt);
            \node[anchor=south] at (\x,\y) {\scriptsize \label};
          }
          \filldraw[blue] (-0.07,0) circle (2pt);
          \filldraw[blue] ( 0.07,0) circle (2pt);
          \node[anchor=north] at (0,-0.05) {\scriptsize $(0,0)$};
        \end{tikzpicture}
        \caption{Meson octet \((1,1)\)}
        \label{fig:meson-octet}
    \end{subfigure}
    \hfill
    \begin{subfigure}[b]{0.32\textwidth}
        \centering
        \begin{tikzpicture}[scale=1.3]
          \draw[->] (-2,0) -- (2,0) node[right] {$I_3$};
          \draw[->] (0,-2.4) -- (0,1.24) node[above] {$Y$};
          \foreach \x/\y/\label in {
            1.5/1/{$(\frac{3}{2},1)$}, 0.5/1/{$(\frac{1}{2},1)$}, -0.5/1/{$(-\frac{1}{2},1)$}, -1.5/1/{$(-\frac{3}{2},1)$},
            1/0/{$(1,0)$}, 0/0/{$(0,0)$}, -1/0/{$(-1,0)$},
            0.5/-1/{$(\frac{1}{2},-1)$}, -0.5/-1/{$(-\frac{1}{2},-1)$},
            0/-2/{$(0,-2)$}}
          {
            \filldraw[blue] (\x,\y) circle (2pt);
            \node[anchor=north] at (\x,\y) {\scriptsize \label};
          }
        \end{tikzpicture}
        \caption{Baryon decuplet \((3,0)\)}
        \label{fig:baryon-decuplet}
    \end{subfigure}
    \hfill
    \begin{subfigure}[b]{0.32\textwidth}
        \centering
        \begin{tikzpicture}[scale=1.3]
          \draw[->] (-2,0) -- (2,0) node[right] {$I_3$};
          \draw[->] (0,-1.2) -- (0,2.4) node[above] {$Y$};
          \foreach \x/\y/\label in {
            0/2/{$(0,2)$},
            0.5/1/{$(\frac{1}{2},1)$}, -0.5/1/{$(-\frac{1}{2},1)$},
            1/0/{$(1,0)$}, 0/0/{$(0,0)$}, -1/0/{$(-1,0)$},
            1.5/-1/{$(\frac{3}{2},-1)$}, 0.5/-1/{$(\frac{1}{2},-1)$}, -0.5/-1/{$(-\frac{1}{2},-1)$}, -1.5/-1/{$(-\frac{3}{2},-1)$}}
          {
            \filldraw[blue] (\x,\y) circle (2pt);
            \node[anchor=north] at (\x,\y) {\scriptsize \label};
          }
        \end{tikzpicture}
        \caption{Anti-baryon decuplet \((0,3)\)}
        \label{fig:antibaryon-decuplet}
    \end{subfigure}
\caption{
Weight diagrams of SU(3) multiplets plotted in the $(Y, I_3)$ plane. In the nested Bethe Ansatz description, each irreducible representation is labeled by a Dynkin label $(p, q)$. For the class of fundamental excitations formed by symmetric combinations of holes, the integers $p$ and $q$ count the number of holes in the Bethe root distributions at each nesting rank: $p$ counts holes in the rank-1 equations and $q$ in the rank-2 equations. For example, (a) the meson octet corresponds to $(1,1)$, with one hole in each level; (b) the baryon decuplet corresponds to $(3,0)$, with three holes in the first level and none in the second; and (c) the antibaryon multiplet corresponds to $(0,3)$. The Bethe Ansatz ground state is the SU(3) singlet $(0,0)$, in which all Bethe roots are real and both levels are fully filled with no holes. This hole-counting interpretation applies only to symmetric excitations. More generally, antisymmetric or mixed-symmetry states—such as those arising in the tensor product decomposition of multiple fundamental excitations—require the inclusion of string solutions, i.e., complex bound states of Bethe roots.
}
    \label{fig:su3-multiplets}
\end{figure*}

We proceed to study variations away from the quantum numbers that we have identified as the ground state. These correspond to the excitations and show that their energy is positive relative to the ground state, allowing us to verify that the identified Bethe Ansatz solution indeed represents the ground state.  These excitations are generated by independently varying the momentum quantum numbers $n_j$ and the two SU(3) color charges $I_\alpha$ and $\mathcal{I}_k$, demonstrating the decoupling of charge and color degrees of freedom.

We shall find that the excitation characterizes three phases as we discussed above on the basis of the renormalization group flow. In the Kondo phase, which we will analyze first,  we shall see that the excitation spectrum consists entirely of bulk excitations, excitations that can propagate across the system.  By contrast, as we shall see, the other phases host both bulk excitations and additional low-energy, non-propagating modes that are localized near the boundary.

\paragraph{Colorless charge excitations:}
The simplest excitations are colorless charge modes that leave the color quantum numbers unchanged. Label the initial quantum numbers $n_j^0$ so that
$$
-2\pi D \le \frac{2\pi}{L}n_j^0 < 0,
$$
and promote one to
$$
n_j' = n_j^0 + \Delta n \ge 0.
$$
The resulting energy change is
$$
\Delta E_c = \frac{2\pi}{L}\Delta n > 0.
$$
Here, $2\pi D$ provides the cutoff that defines the “bottom of the sea” for these charge excitations. Note that while the standard Wilsonian cutoff scheme is imposed on bare (noninteracting) momenta, our cutoff $2\pi D$ is imposed to the fully interacting momentum variables determined by the exact Bethe Ansatz solution. These colorless charge modes are gapless and are commonly called holons. In contrast, the excitations in the color sector are more complex: they arise from distinct solutions of the Bethe Ansatz equations (Eqs.\eqref{bae:rank1-eqn} and \eqref{bae:rank2-eqn}), reflecting the two‐level rapidity structure of the $SU(3)$ model.

{\it{Color excitations}}: these are constructed by varying the color quantum numbers $I_\alpha$ and $\mathcal I_k$ which induces holes in the rank-1 or rank-2 Bethe equations such that the physical excitations are colorless (i.e. the triality $t \equiv (p + 2q)\bmod 3 = 0$). We begin by computing their energies in the presence of the cutoff $D$ and running coupling $b$.  A hole in the rank-1 equation at rapidity $\theta$ shifts the first-level density by

\begin{equation}
\Delta\tilde\rho_1(\omega)
= -\frac{(e^{|\omega|}+1)\cos( \omega \theta)}{2\cosh(\omega)+1},
\end{equation}

and from Eq.~(\ref{engeqn}) its energy is

\begin{equation}
E_\theta
=2D\tan^{-1}\left[\frac{2\mathrm{csch}(2\pi b/3)\cosh(2\pi\theta/3)-\coth(2\pi b/3)}{\sqrt3}\right].
\end{equation}

This excitation carries Dynkin labels $(p,q)=(1,0)$, i.e.\ it lies in the $\mathbf{3}$ (fundamental) representation of $SU(3)$.  A hole in the rank-2 equation at rapidity $\vartheta$ shifts the same density by

\begin{equation}
\Delta\tilde\rho_2(\omega)
= -\frac{e^{|\omega|/2}\cos( \omega \vartheta)}{2\cosh(\omega)+1},
\end{equation}

and its energy is

\begin{equation}
E_\vartheta
=2D\tan^{-1}\left[\frac{2\mathrm{csch}(2\pi b/3)\cosh(2\pi\vartheta/3)+\coth(2\pi b/3)}{\sqrt3}\right].
\end{equation}

This excitation is denoted by Dynkin labels $(p,q)=(0,1)$, i.e.\ it lies in the $\overline{\mathbf{3}}$ (antifundamental) representation.  Although  the energy of rank-1 hole $E_\theta$ and rank-2 hole $E_\vartheta$ differ in the presence of the UV cutoff $D$ as a function of the bulk running coupling constant $b$, both energies become universal in the double-scaling limit
$b\to\infty,\quad D\to\infty$ while holding 
$\bar m = 2De^{-2\pi b/3}$ fixed.  In this limit, each hole takes the relativistic form

\begin{equation}
E(\phi)=\frac{\sqrt3}{2}\bar m\cosh\!\left(\frac{2\pi\phi}{3}\right),
\quad \phi=\theta,\vartheta,
\end{equation}

showing that both rank-1 and rank-2 holes acquire the same physical mass $m=\frac{\sqrt{3}}{2}\bar m$\footnote{We include the factor $\frac{\sqrt{3}}{2}$ up front because, in an $SU(N)$ theory, the mass of a rank-$r$ hole is generally given by
\[
m_r = m\sin\!\left(\frac{\pi r}{N}\right).
\] as shown in Ref.\cite{andrei1980derivation}
}. 

After computing the physical mass of the holes in each rank, we are now in a position to describe the massive fundamental color excitation in the model. Interestingly, the model does not admit isolated single-quark excitations in bulk; only color-neutral hadrons, i.e., bound states of quark and antiquark (mesons) or three-(anti)quark (baryons), can occur as physical excitations.

\paragraph{The meson octet:}
The most fundamental massive color excitation in the model is the bound state of a quark and an antiquark, obtained by introducing one hole into each of the rank-1 and rank-2 Bethe equations. Denote the hole rapidity in the rank-1 sector by $ \theta $ and that in the rank-2 sector by $ \vartheta $ (see Appendix~\ref{sec:BAE-soln}). Its energy is 
$ E(\theta,\vartheta) = m\left(\cosh\frac{2\pi}{3}\theta + \cosh\frac{2\pi}{3}\vartheta\right) $, 
which when the rapidities are sent to infinity $ \theta,\vartheta\to\infty $ attains its minimal value $ E_{\min} = 2m $. This excitation transforms in the Dynkin representation $(1,1)$, i.e.\ the 8-dimensional adjoint multiplet of $SU(3)$ with triality 0. The highest‐weight state has $I_3=\tfrac12$ and $Y_{\max}=1$, and the rest seven states are obtained by successive application of the lowering operators $E_{-\alpha_1}$ and $E_{-\alpha_2}$. These eight states give the usual meson‐octet diagram familiar from QCD, which we explicitly plot in the $I_3$-$Y$ plane in Fig.\ref{fig:meson-octet}.

\paragraph{The meson singlet:}
When one adds a single two-string in each sector (i.e.\ $\lambda\pm\frac{i}{2}$ in the rank-1 equation and $\mu\pm\frac{i}{2}$ in the rank-2 equation) on top of one hole per rank at rapidities $\theta$ (rank-1) and $\vartheta$ (rank-2), the resulting state is the meson singlet with Dynkin labels $(0,0)$(see Appendix~\ref{sec:BAE-soln}). Its energy is given by $E(\theta,\vartheta)=m\left(\cosh\frac{2\pi}{3}\theta+\cosh\frac{2\pi}{3}\vartheta\right)$, which again attains its minimal value $E_{\min}=2m$ when rapidities $\theta,\vartheta\to \infty$. The energy contribution in the thermodynamic limit comes solely from the two holes since the bare energy of the two-strings is exactly canceled by the contribution from the backflow of the roots.

\paragraph{The baryon decuplet:}
If one creates three holes in the rank-1 Bethe equation at rapidities $\theta_i$ $(i=1,2,3)$ (see Appendix~\ref{sec:BAE-soln}), the resulting excitation transforms in the Dynkin representation $(3,0)$ with triality 0. Its energy is  
$
E(\theta_1,\theta_2,\theta_3)
= m\sum_{i=1}^3 \cosh\left(\frac{2\pi}{3}\theta_i\right),
$ 
which as $\theta_i\to\infty$ approaches its minimal value $E_{\min}=3m$. This representation is 10-dimensional, with $Y_{\max}=1$ and $I_3=\frac32$. The remaining nine descendant states are then obtained by successive applications of the two lowering operators, which are shown in the weight diagram in Fig.\ref{fig:baryon-decuplet}.

\paragraph{The anti-baryon decuplet:}
If one creates three holes in the rank-2 Bethe equation at rapidities $\vartheta_i$ $(i=1,2,3)$ (see Appendix~\ref{sec:BAE-soln}), the resulting excitation transforms in the Dynkin representation $(0,3)$. Its energy is  
$E(\vartheta_1,\vartheta_2,\vartheta_3)
= m\sum_{i=1}^3 \cosh\left(\frac{2\pi}{3}\vartheta_i\right)$,  
which as $\vartheta_i\to\infty$ approaches its minimal value $E_{\min}=3m$. This representation is 10-dimensional, with $Y_{\max}=2$ and $I_{3,\max}=0$. The nine descendant states are obtained by applying lowering operators $E_{-\alpha_1}$ and $E_{-\alpha_{2}}$ as shown in the weight diagram in Fig.\ref{fig:antibaryon-decuplet}.

All other higher-order excitations with a larger number of quarks and antiquarks are obtained by adding appropriate numbers of holes in the two-rank equations, such that one gets color-neutral excitations. We refer to Section VII of Ref.~\cite{andrei1983solution} for further discussion on the allowed color excitations.

\subsubsection{Bulk Density of states}
From the Fourier-space Bethe Ansatz density of rank-1 roots in Eq.~\eqref{rank1fs}, the bulk contribution (proportional to $N_q$) in $\lambda$-space becomes
\begin{equation}
\rho_1(\lambda)
=\frac{N_q}{2\sqrt{3}}
\sum_{\upsilon=\pm1}
\frac{1}
     {2\cosh\!\left(\frac{2\pi}{3}(b+\upsilon\lambda)\right)-1},
\end{equation}
such that the bulk density of states in the scaling limit becomes

\begin{equation}
    \mathrm{DOS}_{\mathrm{bulk}}=\left|\frac{\tilde \rho_1^{\mathrm{bulk}}(\lambda)}{E'(\lambda)}\right|=\frac{L}{4\pi}\frac{E}{\sqrt{E^2-m^2}}.
\end{equation}

Note that this form of the DOS is precisely the superconducting density of states, with the characteristic square‐root divergence at the gap edge $E=m$. Moreover, for energies well above the gap ($E\gg m$), it reduces to the constant value $L/(4\pi)$, as expected for a 1+1D non‐interacting Fermi gas.

\subsubsection{The Kondo scale}

Deep in the Kondo phase ($d \gg 1$), where the boundary Kondo physics overwhelms the bulk superconductivity,  apart from the scale of mass gap $m$, the impurity generates a new scale called the Kondo temperature $T_K$. By isolating the impurity parameter $d$ dependent term in the root density of rank-1 in the ground state given by Eq.~\eqref{rank1fs}, using 
 $\mathrm{DOS}_{\mathrm{imp}}(E)=\rho_{\mathrm{imp}}(\lambda)/E'(\lambda)$, one finds
\begin{equation}
\mathrm{DOS}_{\mathrm{imp}}(E)
= \frac{\frac{2 E \cosh\left(\frac{2\pi d}{3}\right) - m}{2\pi m \sqrt{E^2 - m^2}}}
       {
        \left[\cosh\left(\frac{4\pi d}{3}\right)
        + \cosh\left(2\cosh^{-1}\left(\frac{E}{m}\right)\right)\right]}.
\end{equation}

The Kondo scale can then be chosen to be $T_K=\frac{m}{2}e^{\frac{2 \pi  d}{3}}$ such that the ratio of the boundary to bulk per unit length density of states $R(E)=L\frac{\mathrm{DOS}_{\mathrm{imp}}(E)}{\mathrm{DOS}_{\mathrm{bulk}}(E)}$ can be for large $d$ in the familiar Lorentzian form
\begin{equation}
    R(E)=\frac{\sqrt3}{2\pi}\frac{T_K}{E^2+T_K^2}.
\end{equation}

The dynamically generated Kondo temperature $T_K$ governs the dynamics of the impurity in this parametric regime. It is the crossover scale below which the impurity quark becomes completely screened by a collective “cloud” of bulk quarks, forming a many-body singlet. Above $T_K$, the impurity quarks behave like asymptotically free quarks. 

In summary, the Kondo phase is characterized by the complete screening of the impurity below a dynamically generated Kondo scale $T_K$, even in the presence of a gapped, color-superconducting bulk. This screening leads to the vanishing of impurity entropy in the infrared, a hallmark of the Kondo effect. After screening, the low-energy spectrum consists solely of bulk color-singlet excitations, such as mesons and baryons, consistent with confinement. The model naturally reveals two emergent scales: the bulk superconducting gap $m$, and the Kondo crossover scale $T_K$, illustrating how impurity screening can persist within a gapped medium such as superconducting quark matter. A closely related phenomenon was explored in Ref.~\cite{kattel2025edge}, where some of us studied the Kondo effect in a gapped spin chain. While that system shares several key features with the present model, such as impurity screening within a gapped medium and the emergence of a Kondo scale, it also exhibits important differences, most notably the presence of competing magnetic order in the bulk, which is absent in the color-superconducting quark matter considered here.

\subsection{The YSR Phase}
As the parameter $\delta$ increases within the range $\frac{1}{2} < \delta < 2$, the strength of the bulk superconducting order becomes comparable to that of the boundary Kondo scale. In this specific regime, a bound mode forms, localized at the boundary hosting the impurity quark. This boundary-bound mode screens the impurity quark, a mechanism analogous to the screening of surface impurities by Yu-Shiba-Rusinov (YSR) bound states in BCS superconductors~\cite{Yu,Shiba,Rusinov}. However, a key distinction is that while YSR states in traditional BCS systems exist everywhere in the phase space, strong quantum fluctuations in the present context restrict this YSR-like screening phenomenon to this narrow parametric range. Recent work has highlighted the possible realization of such a phase in dense quark matter~\cite{kanazawa2016overscreened,baran2025yu} with impurities. 

The bound mode is described by a unique, purely imaginary solution to the rank-1 Bethe Ansatz equation. This solution, $\lambda_\alpha$, takes the form:
\begin{equation}
\lambda_\alpha = \pm i \left(\frac{1}{2} - \delta\right). \label{eq:lambda_alpha_boundary_string}
\end{equation}
Such solutions arising from Bethe Ansatz equations are characteristic of systems with open boundary conditions and are referred to as boundary-string solutions~\cite{wang1997exact,kattel2024overscreened,kattel2025edge,kattel2024kondo,rylands2020exact,kapustin1996surface,PhysRevB.105.174517}. The nested Bethe Ansatz structure of this problem also permits boundary-string solutions. For the rank-2 equation, when $\delta>1$, such a solution takes the form:
\begin{equation}
    \mu_k = \pm i(1-\delta).\label{eq:mu_k_boundary_string}
\end{equation}
A state where the impurity is screened by a boundary-localized bound mode can then be constructed by adding the boundary-string solution from the rank-1 Bethe Ansatz equations for the range $1/2 < \delta < 1$, and by adding both boundary-string solutions for the range $1 < \delta < 2$ (see Appendix~\ref{sec:BAE-soln} for details).

In this regime, the state formed by all real roots of the rank-1 and rank-2 equation yields a root density of the form:
\begin{align}
    \tilde\rho_1(\omega)&=\frac{2 \left(e^{| \omega | }+1\right) N_q \cos (b \omega )+\frac{\left(1-e^{2 | \omega | }\right)}{e^{\delta  | \omega | }} -e^{\frac{3 |
   \omega | }{2}}+1}{2 (1+2 \cosh (| \omega | )) e^{\frac{| \omega | }{2}}}\\
   \tilde\rho_2(\omega)&=\frac{e^{-\frac{| \omega | }{2}}-e^{| \omega | }-e^{-\delta 
   | \omega | } \left(-1+e^{| \omega | }\right)+2 \cos (b
   \omega ) N_q}{2 (1+2 \cosh (| \omega | ))},
\end{align}
such that the number of rank-1 roots is $M_1=\int \rho_1(\lambda)\mathrm{d}\lambda=\frac{2}{3}N_q$, and the number of rank-2 roots is $M_2=\int \rho_2(\mu)\mathrm{d}\mu=\frac{N_q}{3}$. Consequently, the state $\ket{\psi_1}$ has Dynkin labels (1,0) and a $\mathbb{Z}_3$ triality of 1, which suggests that the impurity quark is unscreened.

Adding the purely imaginary root Eq.~\eqref{eq:lambda_alpha_boundary_string} in the regime $\frac{1}{2}<\delta<1$ and adding both imaginary roots Eq.~\eqref{eq:lambda_alpha_boundary_string} and Eq.~\eqref{eq:mu_k_boundary_string} in the regime $1<\delta<2$, results in a state $\ket{\psi_\delta}$ with Dynkin labels (0,0) with $\mathbb{Z}_3$ triality 0. In this state, the impurity quark is completely screened by the single particle localized bound mode. The change in root density of the rank-1 equation due to the presence of these roots is 
\begin{align}
    \Delta\tilde\rho_1(\omega)=-\frac{e^{-\frac{1}{2} (1-2 \delta ) | \omega | }
   \left(e^{(1-2 \delta ) | \omega | }+1\right)}{2 \left(e^{|
   \omega | }+e^{2 | \omega | }+1\right)}
\end{align}

The energy of the bound mode obtained using Eq.~\eqref{engeqn} and adding its bare energy, which in the scaling limit becomes
\begin{equation}
    E_\delta=-m \sin \left(\frac{\pi}{6} (1+4  \delta
    )\right).
    \label{Eng-bm-eqn}
\end{equation}

\begin{figure}[H]
    \centering
    \includegraphics[width=\linewidth]{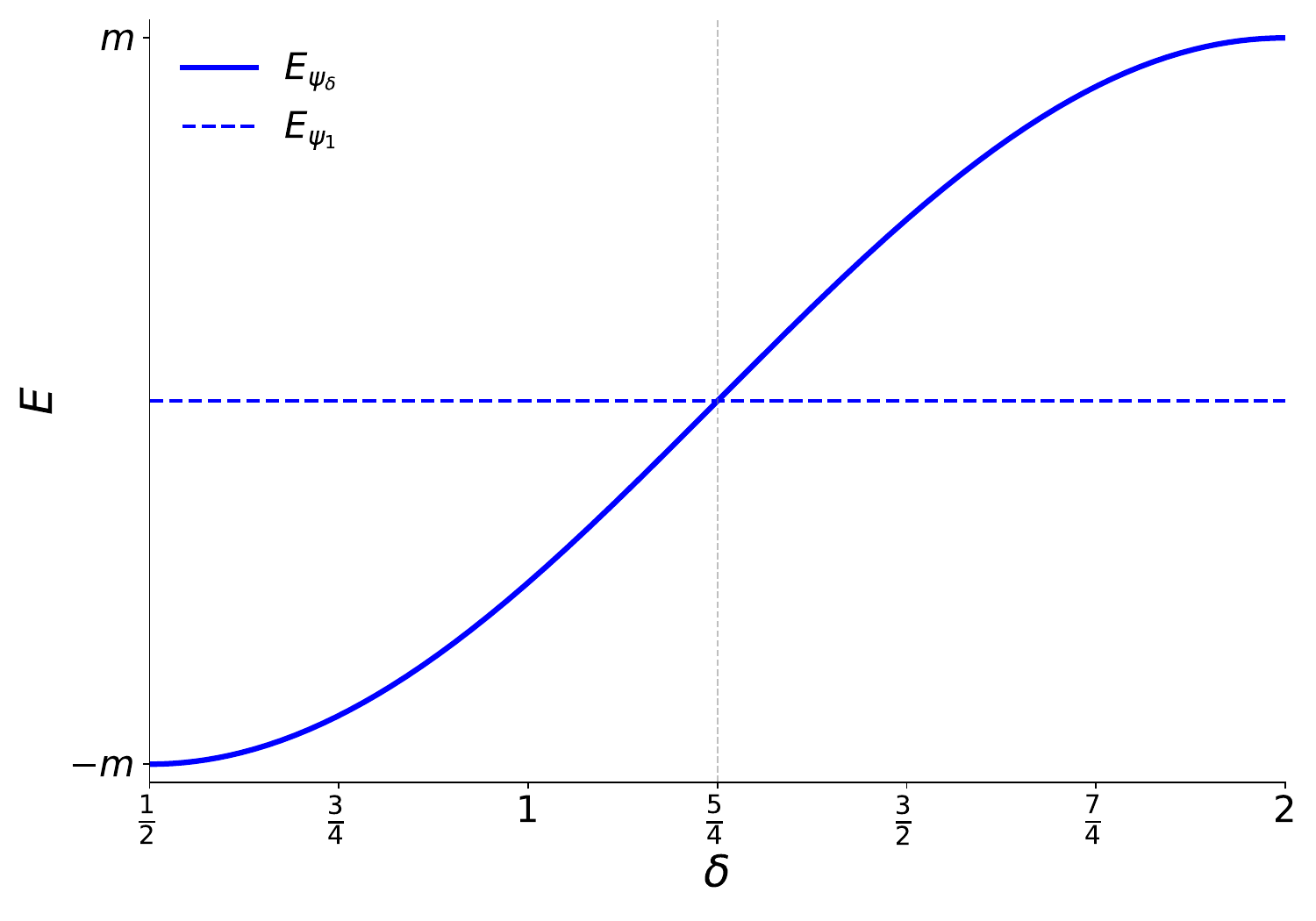}
    \caption{Energy of the boundary bound mode (i.e, the energy difference between state $\ket{\psi_1}$ and state $\ket{\psi_\delta}$) as given by Eq.~\eqref{Eng-bm-eqn}, showing that the mode’s energy is negative for 
$\frac{1}{2}<\delta<\frac{5}{4}$ and becomes positive for $\frac{5}{4}<\delta<2$.}
    \label{fig:Eng-bm-delta}
\end{figure}

Notice that the energy of this root is negative in the range $\frac{1}{2}<\delta<\frac{5}{4}$ and positive in the range $\frac{5}{4}<\delta<2$ as shown in Fig.~\ref{fig:Eng-bm-delta}.

Adding boundary strings to the state $\ket{\psi_1}$ results in the ground state. When $1/2 < \delta < 5/4$, a single-particle bound mode screens the impurity. However, for $5/4 < \delta < 1$, this mid-gap state's energy is higher than that of $\ket{\psi_1}$. This indicates a boundary quantum phase transition in the model at $\delta = 5/4$ due to the level crossing as shown in Fig.~\ref{fig:Eng-bm-delta}. For $\delta < 5/4$, the impurity quark is screened in the ground state, whereas for $\delta > 5/4$, it is unscreened.

All of the bulk excitations described in Sec.~\ref{bulk-excitation} remain valid in this phase. In addition, there are boundary excitations involving the unscreened quark in the reference state $\ket{\psi_{1}}$. For example, if one adds two holes at rapidities $\theta_{1}$ and $\theta_{2}$ into the rank-1 Bethe equation on top of $\ket{\psi_{1}}$, one obtains the state $\ket{\psi_{\theta_{1},\theta_{2}}}$. In this configuration, the unscreened quark, together with the two fundamental holes, carries Dynkin labels $(3,0)$ and triality $0$. Its energy relative to $\ket{\psi_{1}}$ is
\begin{equation}
E_{\psi_{\theta_{1},\theta_{2}}} - E_{\ket{\psi_{1}}}
= m\cosh\left(\frac{2\pi}{3}\theta_{1}\right)
+ m\cosh\left(\frac{2\pi}{3}\theta_{2}\right).
\end{equation}
In the limit $\theta_{1},\theta_{2}\to +\infty$, each $\cosh\left(\frac{2\pi}{3}\theta_{i}\right)\to 1$, so the minimal excitation energy is $2m$. Physically, this boundary excitation describes a baryonic decuplet: the unscreened quark at the boundary and the two massive propagating bulk quarks (arising from the holes in the rank 1 equation) form a $\mathbb{Z}_3$ triality-zero decuplet state.

Similarly, creating a single hole at rapidity $\vartheta$ into the rank-2 Bethe equation on top of $\ket{\psi_{1}}$ produces the state $\ket{\psi_{\vartheta}}$. In this case, the excitation made up of an unscreened quark and the single hole carry Dynkin labels $(1,1)$ and triality $0$. Their energy difference relative to $\ket{\psi_{1}}$ is
\begin{equation}
E_{\psi_{\vartheta}} - E_{\ket{\psi_{1}}}
= m\cosh\left(\frac{2\pi}{3}\vartheta\right),
\end{equation}
which approaches its minimum value $m$ as $\vartheta\to +\infty$. In other words, the unscreened boundary quark, together with one bulk quark, forms the meson octet in the excited spectrum. It is also possible to construct a degenerate singlet state by adding two-string solutions to both the rank 1 and rank 2 equations.

Thus far, all bulk excitations transform in the triality-$0$ (color‐singlet) sector, whereas at the boundary, the decoupling of the impurity quark permits triality‐$1$ states. We now demonstrate that, within the YSR phase and in certain parametric windows, even more exotic boundary excitations emerge.
In the interval $1 < \delta < \frac{3}{2}$, adding to the sea of real root solutions (of both the rank‐$1$ and rank‐$2$ Bethe equations) only the rank‐$1$ boundary string solutions, one obtains a boundary excitation carrying triality~$2$. The total number of rank 1 roots including the one purely imaginary boundary string solution is $M_1=1+\int\rho_1(\lambda)\mathrm{d}\lambda=\frac{1}{3} \left(2 N_b+1\right)$ and the total number of rank 2 roots is $M_2=\int\rho_2(\mu)\mathrm{d}\mu=\frac{1}{3} \left(N_b-1\right)$ (see Appendix Sec.\ref{ysr-detailed}). Thus, the Dynkin label for this state is $(p,q)=(0,1)$. The change in the density of rank 1 roots due to the boundary string solution is
\begin{equation}
    \Delta\tilde\rho_1(\omega)=-\frac{e^{\delta  | \omega | -\frac{3 | \omega | }{2}} \left(e^{(1-2 \delta ) | \omega | }+e^{(3-2 \delta ) | \omega | }+e^{| \omega |
   }+1\right)}{2 (2 \cosh (\omega )+1)},
\end{equation}
such that the total energy of this excitation, including the bare energy and the energy due to the back flow of roots, computed using Eq.\eqref{engeqn} 
which upon taking the double scaling limit $D\to\infty$ and $b\to\infty$ becomes
\begin{equation}
    E_\delta=-m\cos\left(\frac{2\pi}{3}\delta\right),
\end{equation}
upon taking the double scaling limit $D\to\infty$ and $b\to\infty$.

When $1<\delta<\frac32$, one finds that $\frac{1}{2}<E_\delta<m$, so the excitation is strictly mid-gap. This three-dimensional boundary excitation contains an unscreened antiquark rather than an unscreened quark. 

In the parametric regime $\delta>\frac{3}{2}$, adding only the rank 1 boundary string solution and a hole at position $\theta$ in the rank 1 solution leads to another triality 2 excitation with Dynkin label $(p,q)=(0,1)$ that transforms in the antifundamental representation of $\rm SU(3)$. The change in the density of rank 1 roots due to the presence of the hole and the boundary string solution is
\begin{equation}
    \Delta\tilde\rho_1(\omega)=\frac{1}{2} \left(e^{| \omega | }-1\right) e^{\frac{1}{2} (1-2 \delta ) | \omega | }+\frac{\left(e^{| \omega | }+1\right) \cos (\theta 
   | \omega | )}{2 \cosh (| \omega | )+1},
\end{equation}
where the first term is due to the presence of the boundary string solution, and the second term is due to the presence of the hole. The energy of this excitation in the scaling limit is
\begin{equation}
    E_\theta=m\cos\left(\frac{2\pi}{3}\theta\right),
    \label{triality2-eng-g32}
\end{equation}
as the energy of the boundary string vanishes in the thermodynamic limit. 

In summary, in the intermediate YSR phase, realized for $\frac{1}{2} < \delta < 2$, impurity screening is no longer driven by a many-body Kondo effect but instead occurs via the formation of a localized single-particle bound state near the impurity. The energy of this bound state,$|E_\delta|<m$, plays a decisive role in determining the ground state structure. For $\frac{1}{2} < \delta <\frac54$, $E_\delta < 0$ and hence occupying the bound state minimizes the energy, and the system favors a fully screened configuration corresponding to the SU(3) singlet state with Dynkin label $ (p,q) = (0,0) $. Conversely, for the parametric regime, $\frac54<\delta<2$, $ E_\delta > 0 $ and hence the bound state remains unoccupied in the ground state, and the impurity quark exists as an unscreened degree of freedom, yielding a state with $ (p,q) = (1,0) $. The transition between these two regimes occurs via a level crossing at $ E_\delta = 0 $, which occurs at $\delta=\frac54$, marking a first-order quantum phase transition. In addition to the bulk spectrum of color-singlet mesons and baryons, this phase also supports localized boundary excitations, wherein the unscreened impurity quark can bind with a bulk antiquark or two bulk quarks to form a meson-like or baryon-like state, respectively. Moreover, for $\delta>1$, one can realize an exotic triality-$2$ boundary excitation. In the interval $1<\delta<3/2,$ its energy lies strictly within the gap, $E_\delta<m$, whereas for $\delta>3/2$ it asymptotically coincides with the energy of a fundamental hole in the thermodynamic limit.

\section{The Unscreened Phase}
As the RG invariant parameter $\delta$ increases further and takes a value in the range $\delta>2$, the bulk superconducting order overwhelms the Kondo scale, and hence the impurity remains unscreened. In this regime, the beta function of the boundary coupling $\beta(J)$ becomes positive, i.e., the Kondo coupling effectively becomes ferromagnetic, and hence it flows to weak coupling. 

In this regime, the two purely imaginary boundary‐string solutions from Eq.~\eqref{eq:lambda_alpha_boundary_string} and Eq.~\eqref{eq:mu_k_boundary_string} remain valid.  The ground state is $  \ket{\psi_{1}}$,
in which the impurity is unscreened. Adding only the rank 1 boundary string solution leads to triality 2 excitation, discussed above with energy given by Eq.\eqref{triality2-eng-g32}. If one adds both boundary strings to the Bethe‐root configuration of $\ket{\psi_{1}}$, then the rank‐1 root density density are changed by
\begin{equation}
  \Delta\tilde\rho_{1}(\omega)
  = -\frac{1}{2} \left(1-e^{| \omega | }\right) e^{\frac{1}{2}
   (1-2 \delta ) | \omega | }.
\end{equation}
Substituting this into Eq.~\eqref{engeqn} shows that the change in bulk energy is exactly canceled by the bare energy of the two boundary strings, so that 
\[
  E_{\delta} = 0.
\]
The two boundary strings solution cannot be added without adding additional massive holes, as this solution is a  `wide string solution'~\cite{destri1982analysis}. Thus, for $\delta>2$, the impurity quark can never be completely screened at any energy scale (see Appendix~\ref{sec:BAE-soln}).

\section{Conclusion}

In this work, we introduced a novel integrable quantum field theory, which is a toy model that captures the competition between color-superconductivity and the color–Kondo effect in quark matter.  In our model, the quarks are massless in the ultraviolet regime; however, an attractive $\mathrm{SU}(3)$ four‐fermion color density–density interaction between opposite‐chirality quarks drives the bulk coupling $g$ to strong coupling in the infrared, dynamically generating a mass gap $m$ (the color‐superconducting gap) via dynamical transmutation. Unlike QCD, our model contains no dynamical gluon bosons; instead, the attraction that gives rise to quasi-long-range superconducting order arises from a four‐fermion color–density–density interaction in bulk. In the presence of a heavy quark impurity the boundary coupling $J$ between the localized impurity and the bulk color density of light quarks renormalizes depending on the relative magnitudes of $J$ and $g$: for $J\gg g$, the system flows into a Kondo‐screened phase; for $J\sim g$, a Yu–Shiba–Rusinov bound state appears within the gap and screens the impurity; and for $g\gg J$, the impurity remains unscreened at low energy. 

In the Kondo phase, we demonstrated that the impurity becomes completely screened below a dynamically generated Kondo scale $T_K$, despite the presence of a finite superconducting gap $m$. The low-energy spectrum is thus governed entirely by bulk color-singlet excitations like mesons and baryons with no boundary impurity excitations. In contrast to the Kondo phase, where the impurity is fully screened by many-body entanglement with the bulk, the intermediate YSR phase features screening via a localized single-particle bound state near the impurity. Consequently, the impurity can be unscreened by not `occupying' the bound mode, which creates an excitation of energy $E_\delta<m$. Finally, there exists an unscreened phase in which the impurity cannot be completely screened at any energy scale and, therefore, retains its free moment entropy value of $\ln 3$, both in the ultraviolet and infrared limits.

We expect that signatures of the phase dynamics of the system may be observed in AMO experiments of high-spin atoms. In our recent study of the SU(2) case, we showed that the impurity thermodynamics exhibits clear signatures of the phase transition~\cite{kattel2025thermodynamics}. The actual quarks appear in various flavors and correspond to a multi-flavor generalization of our model, which we will study next. In addition, it would be interesting to investigate the thermodynamic properties of the SU(3) model, which we plan to address in forthcoming work.

\bibliographystyle{unsrt}
\bibliography{ref}

\widetext

\appendix
In the Appendices, we present supplementary discussions and detailed derivations of concepts and results that were too extensive to include in the main body. Appendix~\ref{npart-state} constructs the explicit N-particle solution of the Bethe Ansatz equations; Appendix~\ref{sec:int} proves the model’s integrability; Appendix~\ref{sec:BAE-deriv} derives the Bethe Ansatz equations; Appendix~\ref{sec:BAE-soln} presents their solution; and Appendix~\ref{RGeqn-deriv} derives the RG equations for both bulk and boundary couplings within our cutoff scheme, which are non-analytically related to Wilsonian cutoff approaches.

\section{$N_q$-particle eigenstate}\label{npart-state}
In this section, we explain the process of determining the eigenstates of the Hamiltonian operator, denoted $ \hat{H} $. Because the particle number operator $ \hat{N} $ is a conserved quantity, satisfying the commutation relation $ [\hat{H}, \hat{N}] = 0 $, we can identify the eigenstates of $ \hat{H} $ by exploring each subspace with a fixed number of bulk quarks $ N_q $, independently. We initiate this analysis with the single-particle case, $ N_q = 1 $, where the wave function can be represented as a superposition of plane waves as
   \begin{equation}
       \ket{k}=\sum_{a_j,\sigma}\int_{-L}^0 \mathrm{d}x e^{i\sigma k x}A^\sigma_{a_1 a_0}\psi_{\sigma,a_1}^\dagger (x) \ket{0}\otimes \ket{a_0},
   \end{equation}
   where $\ket{0}$ is the vacuum $\psi_{\sigma,a_j}\ket{0}=0$. Here $A^\sigma_{a_j a_0}$ are the amplitudes for the $j^{\mathrm{th}}$ itinerant fermion with chirality index $\sigma$ and color index $a_j$ scattering off the localized impurity carrying color $a_0$. Applying the Hamiltonian to the state $\ket{k}$, Schrodinger's equation fixes these amplitudes in the following form
   \begin{equation}
       A^-_{a_ja_0}={S^{j0}}_{a_ja_0}^{b_jb_0} A^+_{b_jb_0},
   \end{equation}
   where the scattering fermion-impurity bare scattering matrix is given by
   \begin{equation}
       S^{j0}=\frac{i I^{j0}+J \vec \tau^j\cdot \vec T^0}{i I^{j0}-J\vec \tau^j\cdot \vec T^0}=e^{i\phi}\frac{I^{j0}-ic P^{j0}}{1-ic},
\label{imp-part-smat}   \end{equation}
   where we suppressed the color indices, and here
   \begin{equation}
       c=\frac {4 J} {1 - \frac {32 J^2} {9}} \quad\quad \text{and}\quad\quad \phi=-2 \tan^{-1}\left(\frac{4J}{3}\right),
   \end{equation}
   and $I^{j0}_{ab,cd}=\delta_{a,b}\delta_{c,d}$ is the identity matrix, and $P^{j0}_{ab,cd}=\delta_{a,d}\delta_{b,c}$ is the  permutation operator, which is related to the Gell-Mann metrics via
   \begin{equation}
       P^{j0}=\frac{1}{2}\tau^j\otimes \tau^0+\frac{1}{3}I^{j0}.
   \end{equation}
   The identification of the impurity-electron S-matrix $S^{j0}$ completes the construction of the eigenstate $\ket{k}$, which has eigenvalue $E=k$.

\subsection{Two particle sector}
Let us now examine the two-particle sector, where the interaction between particles, characterized by the coupling strength $ g $, as specified in the Hamiltonian given by Eq.~\eqref{modelham}, becomes important. We shall express the wave function as a superposition of plane waves, with the amplitudes varying across regions defined by the relative ordering of the particles
\begin{equation}
\ket{k_i,k_j}=\sum_{\sigma,a}\int_{-L}^0 \mathrm{d}^2x F_{a_ia_ja_0}^{\sigma_i\sigma_j}(x_i,x_j)e^{i k_i \sigma _i x_i+i k_j \sigma _j x_j}\psi_{\sigma_ia_i}^\dagger(x_i)\psi_{\sigma_j a_j}^\dagger(x_j)\ket{0}\otimes \ket{a_0},
\label{2partES}
\end{equation}
where we evaluate a sum across all possible combinations of color and chirality, where the two-particle wave function, represented as $ F_{a_i a_j a_0}^{\sigma_i \sigma_j}(x_i, x_j) $, is formulated using scattering amplitudes. This wave function varies across different regions, depending on the spatial arrangement of the particles:

\[
F_{a_i a_j a_0}^{\sigma_i \sigma_j}(x_i, x_j) = A_{a_i a_j a_0}^{\sigma_i \sigma_j}[ij] \theta(x_j - x_i) + A_{a_i a_j a_0}^{\sigma_i \sigma_j}[ji] \theta(x_i - x_j).
\]

In this expression, $ A_{a_i a_j a_0}^{\sigma_i \sigma_j}[\mathcal{C}] $ denotes the amplitude associated with a particular configuration $ \mathcal{C} $, defined by color $ a_j $, chirality $ \sigma_j $, and the positional ordering of the particles. For example, $ \mathcal{C} = ij $ describes the case where the $ i^{\text{th}} $ particle is positioned to the left of the $ j^{\text{th}} $ particle, while $ \mathcal{C} = ji $ indicates the opposite arrangement.

When the Hamiltonian from Eq.~\eqref{modelham} is applied to the state given in Eq.~\eqref{2partES}, we find that this state is an eigenstate with the energy eigenvalue:

\[
E = k_i + k_j,
\]

provided that the amplitudes satisfy relationships mediated by various scattering matrices, which are elaborated below.

As the rightmost particle scatters off an impurity, its chirality flips. The amplitudes before and after this interaction are connected via the impurity-particle scattering matrix $ S $, as derived in Eq.~\eqref{imp-part-smat}. Suppressing color indices for simplicity, these relationships are:

\begin{align}
A^{\sigma_i -}[ij] &= S^{j0} A^{\sigma_i -}, \\
A^{-\sigma_j}[ji] &= S^{i0} A^{-\sigma_j}.
\end{align}

The exchange interaction introduces two types of scattering matrices, labeled $ S^{ij} $ and $ W^{ij} $. The matrix $ S^{ij} $ specifically relates amplitudes when particles with opposite chirality swap positions:

\begin{align}
A^{+-}[ji] &= S^{ij} A^{+-}[ij], \\
A^{-+}[ij] &= S^{ij} A^{-+}[ji],
\end{align}

where $ S^{ij} $ acts on the color degrees of freedom of the $ i^{\text{th}} $ and $ j^{\text{th}} $ particles. The detailed form of $ S^{ij} $ is provided later.
\begin{equation}
    S^{ij}=e^{i\zeta}\left(\frac{2ib I^{ij}+P^{ij}}{2ib+1} \right),
\end{equation}
where 
\begin{equation}
    b=\frac{9-32 g^2}{72 g} \quad\quad \text{and}\quad\quad \zeta=-2\tan^{-1}\left(\frac{4g}{3}\right).
\end{equation}
Likewise, $W^{ij}$ relates the scattering amplitudes when the particles of the same chirality are exchanged:
\begin{align}
    A^{--}[ji]&=W^{ij}A^{+-}[ij]\\
    A^{++}[ij]&=W^{ij}A^{-+}[ji].
\end{align}

Unlike the scattering matrices $S^{ij}$, the scattering matrices $W^{ij}$ are not dictated by the Hamiltonian as there is no interaction between the particles of the same chirality. Their matrices are fixed via a consistency relation called the reflection equation
\begin{equation}
    S^{i0}S^{ij}S^{j0}W^{ij}=W^{ij}S^{j0}S^{ij}S^{i0},
\end{equation}
which is readily satisfied if we take $W^{ij}=P^{ij}$

Having obtained these various scattering matrices, we have now completed the diagonalization of the Hamiltonian in the two particles sector.

\subsection{$N_q$-particle sector}
Generalization to the $N_q-$bulk quark sector is now fairly straightforward. We construct eigenstates
with eigenvalues 
\begin{equation}
    E = \sum_{j=1}^{N_q} k_j.
\end{equation}
of the form
\begin{equation}
| \{ k_j \} \rangle = \sum_{Q, \vec{a}, \vec{\sigma}} \int \theta(x_Q) A_{\vec{a}}^{\vec{\sigma}}[Q] \prod_{j=1}^{N_q} e^{i \sigma_j k_j x_j} \psi_{a_j \sigma_j}^{\dagger}(x_j) | 0 \rangle \otimes | a_0 \rangle,
\label{NpartES}
\end{equation}
whereas before the sum is taken over all color and chirality configurations that are specified by $\vec a=\{a_1,\cdots,a_{N_q},a_0\}$ and $\vec \sigma=\{\sigma_1,\cdots,\sigma_{N_q},\sigma_0\}$ respectively as well as the different orderings of the particles in the configuration space which correspond to elements of the symmetric group $Q \in S_{N_q}$. Furthermore, $\theta(x_Q)$ represents the Heaviside function, which is nonzero only for a specific ordering. In the $N_q=1, 2$ sectors, the amplitudes $A_{\vec{a}}^{\vec{\sigma}}[Q]$ are interconnected by various $S$ matrices in the same manner as previously described. Notably, the amplitudes differing by the chirality change of the rightmost particle $j$ scattering off the impurity are linked by the impurity-particle $S$ matrix, $S^{j0}$. Those that differ by the order exchange of particles with opposite or identical chiralities are related by $S^{ij}$ and $W^{ij}$, respectively. The validity of this construction is ensured by the $S-$ matrices satisfying the reflection and Yang-Baxter equations:
\begin{align}
W^{jk} W^{ik} W^{ij} &= W^{ij} W^{ik} W^{jk}, \label{Wrel}\\
S^{jk} S^{ik} W^{ij} &= W^{ij} S^{ik} S^{jk}, \label{SWrel}\\
S^{j0} S^{ij} S^{i0} W^{ij} &= W^{ij} S^{i0} S^{ij} S^{j0} \label{YEBrel},
\end{align}
where $W^{ij} = P^{ij}$ with the superscripts indicating the color space of the particles on which the operators act non-trivially. These four relationships are sufficient to construct a consistent $N_q$-particle eigenstate.

\section{Integrability of the model}\label{sec:int}

To show that our model is integrable—meaning we can solve it exactly for its energy levels and wave functions—we proceed step by step, focusing on clarity and detail.

\paragraph{Open Boundary Conditions}:

Consider a system of particles trapped between two reflective walls at $ x = 0 $ and $ x = -L $. We impose {open boundary conditions} to describe how particles behave at these walls. Let $ \psi_{+}(x) $ be the wave function for particles moving right and $ \psi_{-}(x) $ for those moving left. At the boundaries, we require:

\begin{align}
\psi_{+}(0) &= -\psi_{-}(0), \label{eq:bc1} \\
\psi_{+}(-L) &= -\psi_{-}(-L). \label{eq:bc2}
\end{align}

These conditions say that when a particle hits a wall, its right-moving part becomes a left-moving part with a sign change, reflecting it back into the system. 
\paragraph{Quantization of Quasimomenta}:

Now, imagine $ N_q $ particles in this system, each with a quasimomentum $ k_j $ (where $ j = 1, 2, \ldots, N_q $). The wave function is a mix of plane waves, but the boundary conditions limit the possible $ k_j $ values. Applying the condition at $ x = -L $ to the $ N_q $-particle wave function (as in Eq.~\eqref{NpartES}), we get:

\begin{equation}
e^{-2i k_j L} A_{\vec{a}}^{\vec{\sigma}}[Q] = \left( Z_j \right)_{\vec{a} \vec{a}'}^{\vec{\sigma}, \vec{\sigma}'} A_{\vec{a}'}^{\vec{\sigma}'}[Q],\label{eq:quant}
\end{equation}
where,
\begin{itemize}
    \item $ e^{-2i k_j L} $ is the geometric phase a particle picks up traveling from $ x = -L $ to $ x = 0 $ and back—a round trip of distance $ 2L $.
    \item $ A_{\vec{a}}^{\vec{\sigma}}[Q] $ are the amplitudes of the wave function in different states, labeled by quantum numbers $ \vec{a} $ (e.g., color index) and $ \vec{\sigma} $ (e.g., chirality), with $ Q $ as additional particle order labels.
    \item $ Z_j $: The transfer matrix for the $ j $-th particle, which we’ll explore next.
\end{itemize}

This equation ensures the wave function matches up after the particle’s journey, giving us the allowed $ k_j $.

\paragraph{The Transfer Matrix $ Z_j $ and integrability}:

The transfer matrix $ Z_j $ defined as

\begin{equation}
Z_j = W^{j,j-1} \cdot W^{j,1} S^{j,A} S^{j,1} \cdot S^{j,N_q} S^{j,B} W^{j,N} \cdot W^{j,j+1}, \label{eq:transfer}
\end{equation}
 is an operator that encodes the propagation of the $ j^{\mathrm{th}} $ particle through the system, traversing from one boundary to the other and returning to its initial position. As it interacts with the remaining $ N_q-1 $ particles, it accumulates the effects of scattering matrices. The particle begins its journey as a right mover, scattering off the other particles, and upon encountering the impurity, its direction of motion reverses, transitioning to a left mover for the return path. This reversal corresponds to a change in its chirality induced by the impurity scattering event. 

By leveraging the algebraic relations outlined in Eq.~\eqref{Wrel} through Eq.~\eqref{YEBrel}, which include the Yang-Baxter equation and associated consistency conditions, it can be rigorously established that the transfer matrices $ Z_j $ and $ Z_k $ commute for all $ j, k $, satisfying the relation $ [Z_j, Z_k] = 0 $. This commutation property is a fundamental characteristic of integrable systems, indicating that all transfer matrices share a common set of eigenvectors and can thus be diagonalized simultaneously. This feature underpins the exact solvability of the model, enabling a precise determination of its spectral properties.

To find the actual solutions, we diagonalize $ Z_j $ using the {Nested Bethe Ansatz} and functional Bethe Ansatz methods. These methods turn the problem into solvable equations, giving us the eigenvalues of $ Z_j $ and thus the quasimomenta and energies.

\paragraph{Continuous Framework}:
To use this formalism, we embed the bare S-matrices into a continuous framework by defining a spectral parameter-dependent R-matrix, $ R(\lambda) $, and a boundary reflection matrix, $ K(\lambda) $. These are constructed to recover the original bare S-matrices at specific $ \lambda $ values, fully capturing the model's scattering properties. Choosing the $R-$matrix of the $SU(3)$ Heisenberg chain
\begin{equation}
    R^{ij}_{ab}(\lambda)=\frac{i\lambda I^{ij}_{ab}+P^{ij}_{ab}}{i\lambda+1},
    \label{rmatt}
\end{equation}
we obtain the two bulk S-matrices as $W^{ij}=R^{ij}(0)$ and $S^{ij}=R^{ij}(2b)$.   The boundary S-matrix can then be embedded by defining $K^{j0}(\lambda)=R^{j0}(\lambda-d)R^{j0}(\lambda+d)$ such that the bare impurity-quark S-matrix can be written as $S^{j0}=K^{j0}(b)$ when $d=\sqrt{b^2-\frac{2 b}{c}-1}$. The $R-$matrix Eq.~\eqref{rmatt} is a solution of the Yang-Baxter equation
\begin{equation}
    R^{12}(\lambda - \mu) R^{13}(\lambda - \nu) R^{23}(\mu - \nu)
= R^{23}(\mu - \nu) R^{13}(\lambda - \nu) R^{12}(\lambda - \mu)
\end{equation}
and the $R-$ and $K-$matrices together satisfy the reflection equation
\begin{equation}
    R^{ij}(\lambda - \mu) K^{i0}(\lambda) R^{ji}(\lambda + \mu) K^{j0}(\mu)
= K^{j0}(\mu) R^{ij}(\lambda + \mu) K^{i0}(\lambda) R^{ji}(\lambda - \mu)
\end{equation}

The transfer matrix $ Z_1 $ is related to the monodromy matrix $ \Xi_A(\lambda) $ as
\begin{equation}
    Z_1 = \tau(b) = \mathrm{Tr}_A  \Xi_A(b)
\end{equation}
where the double-row monodromy matrix is defined as
\begin{equation}
    \Xi_A(\lambda) = R^{A1}(\lambda + b) \cdots R^{AN_q}(\lambda + b)  R^{A0}(\lambda + d)  R^{A0}(\lambda - d)  R^{AN_q}(\lambda - b) \cdots R^{A1}(\lambda - b).
\end{equation}
Here, $ A $ denotes the auxiliary space, and $ \mathrm{Tr}_A $ represents the trace over the auxiliary space.

Using the properties of the $ R- $matrices (such as unitarity and crossing relations), the Yang-Baxter equation, and the reflection equation, one can prove that the double-row monodromy matrix $ \Xi(\lambda) $ satisfies the reflection equation
\[
R^{ij}(\lambda - \mu)  \Xi_i(\lambda)  R^{ji}(\lambda + \mu)  \Xi_j(\mu)
= \Xi_j(\mu)  R^{ij}(\lambda + \mu)  \Xi_i(\lambda)  R^{ji}(\lambda - \mu),
\]
where $ \Xi_i(\lambda) $ and $ \Xi_j(\mu) $ act on auxiliary spaces $ i $ and $ j $, respectively.

Using the unitarity and crossing symmetry of the $ R $-matrices, and the above reflection equation, one can then prove that the transfer matrix $\tau(\lambda)$
satisfies
\[
[\tau(\lambda), \tau(\mu)] = 0,
\]
for any values of $ \lambda $ and $ \mu $. This commutativity implies the integrability of the model.

\section{Bethe Ansatz equations}\label{sec:BAE-deriv}

All bulk and boundary scattering matrices in our model can be embedded into the $SU(3)$-invariant rational $R$-matrix:
\begin{equation}
    R_{ab}(u) = \frac{i\,u + P_{ab}}{i u + 1},
\end{equation}
where $P_{ab}{}^{ij}{}_{kl} = \delta^i_l \, \delta^j_k$ is the permutation operator on $ \mathbb{C}^3 \otimes \mathbb{C}^3 $.  
This $R$-matrix satisfies the Yang–Baxter equation and is regular i.e. $R(0)=P$. We impose open boundary conditions  and introduce inhomogeneities $b$ in the bulk and $d$ at the boundary impurity.

To diagonalize the transfer matrix, we use the nested $T-Q$ relations and functional Bethe Ansatz method. Note that since the rank of $\mathfrak{su}(3)$ is 2, there are two ranks of Bethe Ansatz equations for the color space and hence there are two sets of Bethe roots and two Baxter $Q$-functions defined by
\begin{align}
    Q^{(1)}(u) &= \prod_{\alpha=1}^{M_1} 
        \left(u - \lambda_{\alpha}^{(1)}\right)\,\left(u + \lambda_{\alpha}^{(1)} + i\right), \\[4pt]
    Q^{(2)}(u) &= \prod_{k=1}^{M_2} 
        \left(u - \mu_{k}^{(2)}\right)\,\left(u + \mu_{k}^{(2)} + 2i\right).
\end{align}
The quasi-momenta $k_j$ of the physical excitations are related to the Bethe roots through
\begin{equation}
    e^{2 i k_j L} = \frac{Q^{(1)}(b-i)}{Q^{(1)}(b)},
\end{equation}
which, after shifting $ \lambda_{\alpha}^{(1)} \mapsto \lambda_{\alpha}^{(1)} - \frac{i}{2}$, becomes
\begin{equation}
    e^{-2 i k_j L}
    =
    \prod_{\alpha=1}^{M_1}\prod_{\upsilon=\pm}
    \frac{\,b + \upsilon \lambda_{\alpha}^{(1)} + \tfrac{i}{2}\,}
         {\,b + \upsilon \lambda_{\alpha}^{(1)} - \tfrac{i}{2}\,}.
\end{equation}

To construct the transfer matrix, we define the forward and reflected monodromy matrices acting on the auxiliary space $A \cong \mathbb{C}^3$ as
\begin{align}
    T_A(u)      &= R_{A,N}(u-b)\cdots R_{A,1}(u-b)\,R_{A,0}(u-d), \\
    \hat{T}_A(u)&= R_{A,0}(u+d)\,R_{A,1}(u+b)\cdots R_{A,N}(u+b),
\end{align}
so that the double-row monodromy matrix is
\begin{equation}
    \Xi_A(u) = T_A(u)\,\hat{T}_A(u),
\end{equation}
and the transfer matrix is defined by
\begin{equation}
    \tau(u) = \mathrm{Tr}_A\!\left[\Xi_A(u)\right].
\end{equation}
The Yang–Baxter equation and reflection algebra ensure the commutativity
\begin{equation}
    [\tau(u),\tau(v)] = 0 \quad \forall\,u,v\in\mathbb{C}.
\end{equation}

The eigenvalues of the transfer matrix satisfy the nested T–Q relation~\cite{wang2015off}
\begin{equation}
    \Lambda(u) =
        a(u)\,\frac{Q^{(1)}(u-i)}{Q^{(1)}(u)}
        +
       d(u) \frac{Q^{(1)}(u+i)\,Q^{(2)}(u-i)}{Q^{(1)}(u)\,Q^{(2)}(u)}
        +
       d(u)\,\frac{Q^{(2)}(u+i)}{Q^{(2)}(u)},
\end{equation}
with scalar prefactors
\begin{align}
    a(u) &= (u+b+i)^N (u-b+i)^N (u+d+i)(u-d+i)\\
    d(u) &= a(-u-i)
\end{align}

Requiring analyticity of $\Lambda(u)$ at the simple poles 
$u = \lambda_{\alpha}^{(1)} $ and 
$u = \mu_{k}^{(2)} $,
and shifting to the symmetric variables
\begin{equation}
    {\lambda}_{\alpha}^{(1)} := \lambda_{\alpha}^{(1)} - \tfrac{i}{2},
    \qquad
    {\mu}_{k}^{(2)} := \mu_{k}^{(2)} - i,
\end{equation}
yields the Bethe Ansatz equations in their final symmetric form.

\paragraph{Rank 1 Bethe equations.}
For the roots $\{\lambda_{\alpha}^{(1)}\}_{\alpha=1}^{M_1}$,
\begin{equation}
\begin{aligned}
\prod_{\upsilon=\pm}
\Biggl[
  \left(
    \frac{\lambda_{\alpha}^{(1)} + \upsilon b - \tfrac{i}{2}}
         {\lambda_{\alpha}^{(1)} + \upsilon b + \tfrac{i}{2}}
  \right)^{\!N}
  \frac{\lambda_{\alpha}^{(1)} + \upsilon d - \tfrac{i}{2}}
       {\lambda_{\alpha}^{(1)} + \upsilon d + \tfrac{i}{2}}
  \prod_{k=1}^{M_2}
    \frac{\lambda_{\alpha}^{(1)} + \upsilon \mu_{k}^{(2)} - \tfrac{i}{2}}
         {\lambda_{\alpha}^{(1)} + \upsilon \mu_{k}^{(2)} + \tfrac{i}{2}}
\Biggr]
&=
\prod_{\substack{\beta=1\\\beta\neq\alpha}}^{M_1}
\prod_{\upsilon=\pm}
  \frac{\lambda_{\alpha}^{(1)} + \upsilon \lambda_{\beta}^{(1)} - i}
       {\lambda_{\alpha}^{(1)} + \upsilon \lambda_{\beta}^{(1)} + i},
\qquad
\alpha = 1,\dots,M_1.
\end{aligned}
\end{equation}

\paragraph{Rank 2 Bethe equations.}
For the roots $\{\mu_{k}^{(2)}\}_{k=1}^{M_2}$,
\begin{equation}
\begin{aligned}
\prod_{\alpha=1}^{M_1}
\prod_{\upsilon=\pm}
  \frac{\mu_{k}^{(2)} + \upsilon \lambda_{\alpha}^{(1)} - \tfrac{i}{2}}
       {\mu_{k}^{(2)} + \upsilon \lambda_{\alpha}^{(1)} + \tfrac{i}{2}}
&=
\prod_{\substack{\ell=1\\\ell\neq k}}^{M_2}
\prod_{\upsilon=\pm}
  \frac{\mu_{k}^{(2)} + \upsilon \mu_{\ell}^{(2)} - i}
       {\mu_{k}^{(2)} + \upsilon \mu_{\ell}^{(2)} + i},
\qquad
k = 1,\dots,M_2.
\end{aligned}
\end{equation}

In the main text and in the following sections, we removed the subscript $()^{(j)}$ from the roots as the use of $\lambda_\alpha$ and $\mu_k$ is enough to distinguish the two sets of the roots. 

\section{Detailed Solution of BAE}\label{sec:BAE-soln}

In this subsection, we shall present detailed solution of the Bethe Ansatz equation in all three boundary phases. 

\subsection{Kondo phase}
Let us consider the case when $d \in \mathbb{R}$. By analytic continuation, all results remain valid for purely imaginary $d = i\delta$ with $0 < \delta < \frac{1}{2}$. Taking the $\log$ on both sides of each of the three equations, we obtain
\begin{equation}
    k_j=\frac{\pi n_j}{L} + \frac{1}{L} \sum_{\alpha=1}^{M_1} \left[ \tan^{-1}(2(b + \lambda_\alpha)) + \tan^{-1}(2(b - \lambda_\alpha)) \right]
\end{equation}
such that summing over $j$ we obtain the energy
\begin{equation}
    E = \sum_{j} k_j = \sum_{j} \frac{\pi n_j}{L} + D \sum_{\alpha=1}^{M_1} \left[ \tan^{-1}(2(b + \lambda_\alpha)) + \tan^{-1}(2(b - \lambda_\alpha)) \right],
\end{equation}
where $n_j$ are the integers, which are charge quantum numbers. 

Moreover, the logarithm of the two ranks of Bethe Ansatz equations read
\begin{align}
    &\sum_{\upsilon=\pm}\left(N_q \tan^{-1}(2 (\lambda_\alpha+ \upsilon b))+\tan^{-1}(2 (\lambda_\alpha+ \upsilon d
   )) +\sum_{k=1}^{M_2}\tan^{-1}(2(\lambda_\alpha+\upsilon\mu_k))\right)+\tan^{-1}(2\lambda_\alpha)\nonumber\\
   &\quad\quad\quad=\pi I_\alpha +\sum_{\upsilon=\pm}\sum_{\beta}\tan^{-1}(\lambda_\alpha+\upsilon\lambda_\beta)
   \label{rank1EQN-app}
\end{align}

and

\begin{equation}
    \tan^{-1}(2\mu_k)+\sum_{\upsilon=\pm} \sum_{\alpha=1}^{M_1}\tan^{-1}(2(\mu_k+\upsilon \lambda_\alpha))=\sum_{\upsilon=\pm}\sum_{\ell=1}^{M_2}\tan^{-1}(\mu_k+\upsilon\mu_{\ell})+ \pi \mathcal{I}_{k}
    \label{rank2EQN-app}
\end{equation}
where $I_\alpha$ and $\mathcal{I}_k$ are integers that represent the color quantum numbers. 

For simplicity, we consider the case where the total number of quarks is a multiple of 3. That is, we take
\[
N = N_q + 1, \quad \text{with } N \equiv 0 \mod 3,
\]
where $N_q$ denotes the number of quarks in the bulk, and the additional $1$ corresponds to a single impurity.

To analyze the coupled equations Eq.~\eqref{rank1EQN-app} and Eq.~\eqref{rank2EQN-app} in the thermodynamic limit, we introduce the densities of roots as follows

\[
\rho_1(\lambda_j)=\frac{1}{\lambda_{j+1}-\lambda_j},
\qquad
\rho_2(\mu_k)=\frac{1}{\mu_{k+1}-\mu_k}.
\]

such that we can convert all the sums over roots to integral using the following identities
\[
\sum_{\beta=1}^{M_1}F(\lambda_\beta)\to\int_{-\infty}^\infty F(\lambda')\rho_1(\lambda')d\lambda',
\quad
\sum_{k=1}^{M_2}G(\mu_k)\to\int_{-\infty}^\infty G(\mu')\rho_2(\mu')d\mu'.
\]

Differentiating with respect to $\lambda_\alpha$ (and similarly $\mu_k$) turns each
$\tan^{-1}(x)$ into the Lorentzian $\frac{d}{dx}\tan^{-1}(x)=\frac{1}{1+x^2}$,
and the two equations become
\begin{align}
2\rho_1(\lambda)&= f(\lambda)+\sum_{\upsilon=\pm}\left(\int K^{\{2\}} (\lambda+\upsilon\mu)\rho_2(\mu)\mathrm{d}\mu-\int K^{\{1\}}(\lambda+\upsilon\lambda')\rho_1(\lambda')\mathrm{d}\lambda'\right)-\delta(\lambda),\\
2\rho_2(\mu)&= a_{\frac{1}{2}}(\mu)+\sum_{\upsilon=\pm}\left(\int K^{\{2\}}(\mu+\upsilon \lambda)\rho_1(\lambda)\mathrm{d}\lambda - \int K^{\{1\}}(\mu+\upsilon \mu')\mathrm{d}\mu'\right)-\delta(\mu)
\end{align}
where the two delta functions are added to remove the two roots $\lambda_\alpha=0$ and $\mu_k=0$ which leads to un-normalizable solutions. Moreover, here
\begin{align}
&f(\lambda)=a_{\frac{1}{2}}(\lambda)+\sum_{\upsilon=\pm}\left(N_q a_{\frac{1}{2}}(\lambda+\upsilon b) +a_{\frac{1}{2}}(\lambda+\upsilon d)\right)\\
&a_\gamma(\lambda)= \frac{1}{\pi}\frac{\gamma}{\lambda^2+\gamma^2}\\
&K^{\{n\}}(\rho)=a_{\frac{1}{n}}(\rho)
\end{align}

Now, we can solve these two equations in Fourier space as
\begin{align}
    2\tilde{\rho}_{1}(\omega)&=\tilde f (\omega)+ 2\tilde{K}^{\{2\}}(\omega)\tilde{\rho}_2(\omega)-2\tilde{K}^{\{1\}}(\omega)\tilde{\rho}_1(\omega)-1\\
     2\tilde{\rho}_{2}(\omega)&=\tilde{a}_{\frac{1}{2}}(\omega)+2\tilde{K}^{\{2\}}(\omega)\tilde{\rho}_1(\omega)-2\tilde{K}^{\{1\}}(\omega)\tilde{\rho}_2(\omega)-1
\end{align}
such that the coupled algebraic equation leads to the solution of the form
\begin{align}
\tilde{\rho}_1(\omega)&=\frac{4 \cosh \left(\frac{\omega }{2}\right) \left(N_q \cos (b \omega )+\cos (d \omega )\right)+e^{-\frac{| \omega
   | }{2}}-e^{| \omega | }}{2 (2 \cosh (| \omega | )+1)}\\
   \tilde{\rho}_2(\omega)&=\frac{e^{-\frac{| \omega | }{2}}-e^{| \omega | }+2 N_q \cos
   (b \omega )+2 \cos (d \omega )}{2 (2 \cosh (| \omega |
   )+1)}
\end{align}

Thus in the ground state, there are
\begin{equation}
    M_1=\int \rho_1(\lambda)\mathrm{d}\lambda=\tilde\rho_1(0)=\frac{2}{3} \left(N_q+1\right)
\end{equation} 
number of roots of the first rank of equations, and there are
\begin{equation}
    M_2=\int \rho_2(\mu)\mathrm{d}\mu=\tilde\rho_2(0)=\frac{1}{3} \left(N_q+1\right)
\end{equation}

Recall that the irreducible representation of $\mathrm{SU}(3)$ is labeled by two Dynkin labels $(\mu_1,\mu_2)$. These labels are obtained from the Bethe Ansatz solution for each state as follows
\begin{align}
    p&=N_q+1-2M_1+M_2\\
    q&=M_1-2M_2,
\end{align}
such that for the ground state, we obtain
\begin{equation}
    p=(N_q+1)-2\frac{2}{3}(N_q+1)+\frac{1}{3}(N_q+1)=0 \quad{\rm and}\quad q=\frac{2}{3}(N_q+1)-2 \frac{1}{3}(N_q+1)=0.
\end{equation}
This shows that the ground state is a many-body $SU(3)$ singlet, and the itinerant bulk quarks completely screen the impurity quark by forming a many-body Kondo cloud.

The fundamental excitations can be constructed by making holes in the roots of either rank. First, by making holes at rank-1 roots, we obtain the change due to each such hole is
\begin{equation}
    \Delta \tilde{\rho}(\theta_1)=-\frac{\left(e^{| \omega | }+1\right) \cos
   \left(\theta_1  \omega \right)}{2 \cosh (\omega
   )+1}
\end{equation}

The energy of the hole computed using Eq.~\eqref{engeqn} becomes
\begin{equation}
E_\theta
=2D\tan^{-1}\left[\frac{2\mathrm{csch}(2\pi b/3)\cosh(2\pi\theta/3)-\coth(2\pi b/3)}{\sqrt3}\right].
\end{equation}
which upon taking the scaling limit $D\to \infty $ and $b\to \infty$ becomes $E_\theta=m\cosh\left(\frac{2\pi}{3}\theta\right)$.

Likewise, the change in density of the rank-1 roots due to the presence of a single hole at the second rank is 
\begin{equation}
    \Delta \tilde{\rho}(\vartheta)=-\frac{e^{\frac{| \omega | }{2}} \cos ( \omega \vartheta )}{2
   \cosh (\omega )+1}
\end{equation}
such that the energy of this hole is 
\begin{equation}
E_\vartheta
=2D\tan^{-1}\left[\frac{2\mathrm{csch}(2\pi b/3)\cosh(2\pi\vartheta/3)+\coth(2\pi b/3)}{\sqrt3}\right],
\end{equation}
which upon taking the scaling limit $D\to \infty $ and $b\to \infty$ becomes $E_\vartheta=m\cosh\left(\frac{2\pi}{3}\vartheta\right)$.

\subsubsection{Fundamental excitations}\label{fund-ext-app}
We briefly discuss the construction of the massive chargeless color fundamental excitations. 

\paragraph{The meson octet:}
Adding a hole in the rank-1 equation at position $\theta$ and a hole in the rank-2 equation at position $\vartheta$, we obtain the solutions of the Bethe Ansatz equations of the form

\begin{align}
    2\tilde{\rho}_{1}(\omega)&=\tilde f (\omega)+ 2\tilde{K}^{\{2\}}(\omega)\tilde{\rho}_2(\omega)-2\tilde{K}^{\{1\}}(\omega)\tilde{\rho}_1(\omega)-1-2\cos(\omega\theta)\\
     2\tilde{\rho}_{2}(\omega)&=\tilde{a}_{\frac{1}{2}}(\omega)+2\tilde{K}^{\{2\}}(\omega)\tilde{\rho}_1(\omega)-2\tilde{K}^{\{1\}}(\omega)\tilde{\rho}_2(\omega)-1-2\cos(\omega\vartheta)
\end{align}
which leads to the solution of the form
\begin{align}
    \tilde\rho_1(\omega)&=\frac{4 \cosh \left(\frac{\omega }{2}\right) N_q \cos (b \omega )+2 e^{\frac{| \omega | }{2}} \left(-2 \cosh
   \left(\frac{| \omega | }{2}\right) \cos (\theta  \omega )+\cos (d \omega )-\cos ( \omega \vartheta )\right)+2 e^{-\frac{| \omega | }{2}}
   \cos (d \omega )+e^{-\frac{| \omega | }{2}}-e^{| \omega | }}{4 \cosh (| \omega | )+2}\\
    \tilde\rho_2(\omega)&=\frac{-2 e^{\frac{| \omega | }{2}} \cos (\theta  \omega )-2 \left(e^{| \omega | }+1\right) \cos ( \omega \vartheta )+e^{-\frac{| \omega |
   }{2}}-e^{| \omega | }+2 N_q \cos (b \omega )+2 \cos (d \omega )}{4 \cosh (| \omega | )+2}.
\end{align}

The number of roots of rank-1 in this state is $M_1=\int\rho_1(\lambda)\mathrm{d}\lambda=\frac{1}{3} \left(2 N_q-1\right)$ and the number of rank-2 roots is $M_2=\int\rho_2(\mu)\mathrm{d}\mu=\frac{1}{3} \left(N_q-2\right)$. Thus, the Dynkin labels of this state are $(p,q)=(1,1)$, and triality is 0. This state represents the meson octet, the bound state between the quark and antiquark. 

The total change in the rank-1 root density due to the presence of these two holes is 
\begin{equation}
    \Delta\tilde{\rho}_1(\omega)=-\frac{\left(e^{| \omega | }+1\right) \cos (\theta  \omega )}{2 \cosh (| \omega | )+1}-\frac{e^{\frac{| \omega | }{2}} \cos (\vartheta 
   \omega )}{2 \cosh (| \omega | )+1}
\end{equation}

Such that using Eq.~\eqref{engeqn}, we find that the energy of the state in the scaling limit is
\begin{equation}
    E_{\theta,\vartheta}=m\left(\cosh\left(\frac{2\pi}{3}\theta\right)+\cosh\left(\frac{2\pi}{3}\vartheta\right)\right),
\end{equation}
which takes the minimum value $E^{\rm min}_{\theta,\vartheta}=2m$ as both the rapidities are sent to infinity.

\begin{figure}[H]
    \centering
\begin{tikzpicture}[scale=1.5]
  \draw[->] (-1.25,0) -- (1.25,0) node[right] {$I_3$};
  \draw[->] (0,-1.25) -- (0,1.5) node[above] {$Y$};

  \foreach \x/\y/\label in {
    1/0/{$(1,0)$}, -1/0/{$(-1,0)$},
    0.5/1/{$(\frac12,1)$}, -0.5/1/{$(-\frac12,1)$},
    0.5/-1/{$(\frac12,-1)$}, -0.5/-1/{$(-\frac12,-1)$}}
  {
    \filldraw[blue] (\x,\y) circle (2pt);
    \node[anchor=south] at (\x,\y) {\scriptsize \label};
  }

  \filldraw[blue] (-0.07,0) circle (2pt);
  \filldraw[blue] ( 0.07,0) circle (2pt);
  \node[anchor=north] at (0,-0.05) {\scriptsize $(0,0)$};
\end{tikzpicture}
\caption{Weight diagram of the $SU(3)$ adjoint $(1,1)$ octet of quark and antiquark bound states plotted in the hypercharge $Y$ versus isospin $I_3$ plane.}
    \label{fig:meson-octet1}
\end{figure}
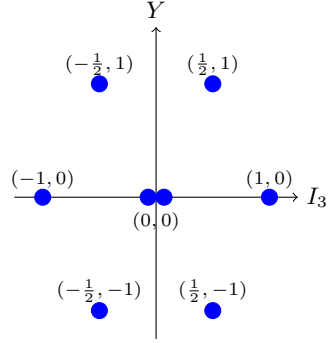

This excitation, labeled by its Dynkin labels $(p,q)=(1,1)$, is 8-dimensional. The highest-weight state has $Y_{\rm max}=1$ and $I_3=\frac{3}{2}$ and all descendant states are obtained by successive application of the lowering operators $E_{-\alpha_1}$ and $E_{-\alpha_2}$ such that the eight states can be plotted in the $Y-I_3$ plane as shown in Fig.\ref{fig:meson-octet1}. 

\paragraph{The meson singlet:}
The meson singlet can then be obtained by adding a two-string solution of the form $\lambda_\alpha=\bar\lambda\pm \frac{i}{2}$ in rank-1 equation and a two-string solution of the form $\mu_k=\bar\mu\pm \frac{i}{2}$ in the second rank equation on top of the holes at position $\theta$ in rank-1 equation and $\vartheta$ in rank-2 equation leads to the Bethe Ansatz equation in Fourier space of the form

\begin{align}
    2\tilde{\rho}_{1}(\omega)&=\tilde f (\omega)+ 2\tilde{K}^{\{2\}}(\omega)\tilde{\rho}_2(\omega)-2\tilde{K}^{\{1\}}(\omega)\tilde{\rho}_1(\omega)-1-2\cos(\omega\theta)+2\cos(\omega\bar\mu)\tilde{a}_1(\omega)-2\cos(\omega\bar\lambda)(\tilde{a}_\frac12(\omega)+\tilde{a}_\frac32(\omega))\\
     2\tilde{\rho}_{2}(\omega)&=\tilde{a}_{\frac{1}{2}}(\omega)+2\tilde{K}^{\{2\}}(\omega)\tilde{\rho}_1(\omega)-2\tilde{K}^{\{1\}}(\omega)\tilde{\rho}_2(\omega)-1-2\cos(\omega\vartheta)+2\cos(\omega\bar\lambda)\tilde{a}_1(\omega)-2\cos(\omega\bar\mu)(\tilde{a}_\frac12(\omega)+\tilde{a}_\frac32(\omega)),
\end{align}

such that the density of roots becomes
\begin{align}
    \tilde\rho_1(\omega)&=\frac{4 \cosh \left(\frac{| \omega | }{2}\right) \left(-\cos \left(\omega  \bar{\lambda }\right)-e^{\frac{| \omega | }{2}} \cos
   (\theta  \omega )+N_q \cos (b \omega )+\cos (d \omega )\right)-2 \frac{ \cos \left(\omega  \bar{\lambda
   }\right)}{e^{\frac{3 | \omega | }{2}}}-2 e^{\frac{| \omega | }{2}} \cos ( \omega \vartheta )+e^{-\frac{| \omega | }{2}}-e^{| \omega | }}{2 (2 \cosh (| \omega | )+1)}
\end{align}
and
\begin{align}
    \tilde{\rho}_2(\omega)=\frac{e^{-\frac{| \omega | }{2}}-e^{| \omega | }-2 e^{\frac{| \omega | }{2}} \cos (\theta  \omega )-2 \frac{   \cos \left(\omega  \bar{\mu }\right)}{e^{\frac{1}{2} (3 | \omega | )}}-4 \left(e^{\frac{| \omega | }{2}} \cos ( \omega \vartheta )+\cos \left(\omega  \bar{\mu
   }\right)\right) \cosh \left(\frac{| \omega | }{2}\right)+2 \left(\cos (d \omega )+\cos (b \omega ) N_q\right)}{2 (1+2 \cosh (| \omega
   | ))}
\end{align}

Such that the total number of rank-1 roots in this state is
\begin{equation}
    M_1=2+\int\rho_1(\lambda)\mathrm{d}\lambda=2+\tilde\rho_1(0)=2+\frac{2}{3} \left(N_q-2\right)=\frac{2}{3} \left(N_q+1\right),
\end{equation}
and the total number of rank-2 roots is
\begin{equation}
    M_2=2+\int\rho_2(\mu)\mathrm{d}\mu=2+\tilde\rho_2(0)=2+\frac{1}{3} \left(N_q-5\right)=\frac{1}{3} \left(N_q+1\right),
\end{equation}
such that the Dynkin labels of this state are (0,0). Which shows that it is a singlet state with $\mathbb{Z}_3$ triality 0. 

The change in the rank-1 root density due to the presence of the two holes and the two two-string solutions is 
\begin{equation}
    \Delta\tilde{\rho}_1(\omega)=-\frac{\left(e^{| \omega | }+1\right) \cos (\theta  \omega )}{2 \cosh (| \omega | )+1}-\frac{e^{\frac{| \omega | }{2}} \cos (\vartheta 
   \omega )}{2 \cosh (| \omega | )+1}-e^{-\frac{| \omega | }{2}} \cos \left(\omega  \bar{\lambda }\right).
\end{equation}

Using Eq.~\eqref{engeqn} and adding the bare energy of the string solutions, we find that the total energy of this state in the scaling limit is
\begin{equation}
    E_{\theta,\vartheta}=m\left(\cosh\left(\frac{2\pi}{3}\theta\right)+\cosh\left(\frac{2\pi}{3}\vartheta\right)\right),
\end{equation}
which takes the minimum value $E^{\rm min}_{\theta,\vartheta}=2m$ as both the rapidities are sent to infinity. In the thermodynamic limit, the energy contribution from the two‐string solution vanishes.  In other words,
\begin{equation}
E_{\overline{\lambda}}
=
D \tan^{-1}\left(\frac{2 b}{\overline{\lambda}^2 - b^2 + 1}\right)
- iD \int \frac{e^{-|\omega| - ib\omega}\cos\left(\omega\overline{\lambda}\right)}{\omega} d\omega
= 0
\end{equation}
where the first term is the bare energy of the string and the second term is the contribution from the back-flow of the roots.

\paragraph{The baryon decuplet:}
Adding three holes in the rank-1 equation at positions $\theta_1,\theta_2$ and $\theta_3$ leads to the Bethe Ansatz in the Fourier space of the form
\begin{align}
    2\tilde{\rho}_{1}(\omega)&=\tilde f (\omega)+ 2\tilde{K}^{\{2\}}(\omega)\tilde{\rho}_2(\omega)-2\tilde{K}^{\{1\}}(\omega)\tilde{\rho}_1(\omega)-1-2\sum_{i=1}^3\cos(\omega\theta_i)\\
     2\tilde{\rho}_{2}(\omega)&=\tilde{a}_{\frac{1}{2}}(\omega)+2\tilde{K}^{\{2\}}(\omega)\tilde{\rho}_1(\omega)-2\tilde{K}^{\{1\}}(\omega)\tilde{\rho}_2(\omega)-1
\end{align}
The solution to which is immediate
\begin{align}
    \tilde\rho_1(\omega)&=\frac{e^{-\frac{| \omega | }{2}}-e^{| \omega | }+4 \cosh \left(\frac{| \omega | }{2}\right) \left(\cos (d \omega )-e^{\frac{| \omega |
   }{2}} \left(\sum_{i=1}^3\cos \left(\omega  \theta_i\right)\right)+\cos
   (b \omega ) N_q\right)}{2 (1+2 \cosh (| \omega | ))}\\
   \tilde\rho_2(\omega)&=\frac{e^{-\frac{| \omega | }{2}}-e^{| \omega | }-2 e^{\frac{| \omega | }{2}} \left(\sum_{i=1}^3\cos \left(\omega  \theta_i\right)\right)+2 \left(\cos (d \omega )+\cos (b \omega ) N_q\right)}{2 (1+2 \cosh (|
   \omega | ))},
\end{align}
such that the number of rank-1 roots is 
\begin{equation}
    M_1=\int\rho_1(\lambda)\mathrm{d}\lambda=\frac{2}{3} \left(N_q-2\right),
\end{equation}
and the number of rank-2 roots is
\begin{equation}
    M_2=\int\rho_2(\mu)\mathrm{d}\mu=\frac{1}{3} \left(N_q-2\right)
\end{equation}

In this case, the Dynkin labels $(p,q) = (3,0)$ correspond to a 10-dimensional, $\mathbb{Z}_3$ triality-0 multiplet. Introducing three holes at rapidities $\theta_1,\theta_2,\theta_3$ changes the rank-1 root density by
\begin{equation}
\Delta \tilde{\rho}(\theta_1,\theta_2,\theta_3)
=
-\frac{(e^{|\omega|}+1)\sum_{i=1}^3 \cos(\theta_i  \omega)}
{2 \cosh(\omega)+1}.
\end{equation}
Using Eq.~\eqref{engeqn}, the energy of this excitation is
\begin{equation}
E_{\theta_1,\theta_2,\theta_3}
=
\sum_{i=1}^3 m \cosh\left(\frac{2\pi}{3} \theta_i\right),
\end{equation}
which attains its minimum value $3m$ as $\theta_i \to \infty$ for all $i=1,2,3$.

\begin{figure}[H]
    \centering
\begin{tikzpicture}[scale=1.5]
  \draw[->] (-2,0) -- (2,0) node[right] {$I_3$};
  \draw[->] (0,-2.4) -- (0,1.24) node[above] {$Y$};
  \foreach \x/\y/\label in {
    1.5/1/{$(\frac{3}{2},1)$}, 0.5/1/{$(\frac{1}{2},1)$}, -0.5/1/{$(-\frac{1}{2},1)$}, -1.5/1/{$(-\frac{3}{2},1)$},
    1/0/{$(1,0)$}, 0/0/{$(0,0)$}, -1/0/{$(-1,0)$},
    0.5/-1/{$(\frac{1}{2},-1)$}, -0.5/-1/{$(-\frac{1}{2},-1)$},
    0/-2/{$(0,-2)$}}
    {
      \filldraw[blue] (\x,\y) circle (2pt);
      \node[anchor=north] at (\x,\y) {\scriptsize \label};
    }
\end{tikzpicture}
    \caption{Weight diagram of the $SU(3)$ decuplet $(3,0)$ of baryons plotted in the hypercharge $Y$ versus isospin $I_3$ plane.}
    \label{fig:Baryon-Decuplet1}
\end{figure}
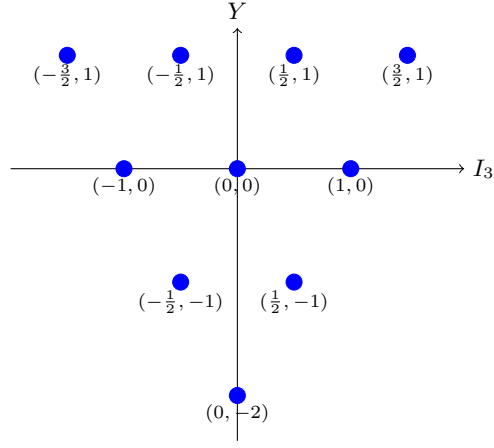

This excitation is 10-dimensional, labeled by its Dynkin labels $(p,q)=(3,0)$. The highest weight state carries $Y_{\rm max}=1$ and $I_3=\frac{3}{2}$. Remaining nine descendants obtained by applying successively the two operators $E_{-\alpha_1}$ and $E_{-\alpha_2}$ are shown in Fig.\ref{fig:Baryon-Decuplet1}.

 The highest-weight state and its nine descendants are plotted in the $Y$–$I_3$ plane in Fig.~\ref{fig:Baryon-Decuplet1}.
\paragraph{The antibaryon decuplet}
Introducing three holes into the rank-1 Bethe equations at rapidities $\vartheta_{1},\vartheta_{2},\vartheta_{3}$ yields, in Fourier space, the following form of the Bethe Ansatz:

\begin{align}
    2\tilde{\rho}_{1}(\omega)&=\tilde f (\omega)+ 2\tilde{K}^{\{2\}}(\omega)\tilde{\rho}_2(\omega)-2\tilde{K}^{\{1\}}(\omega)\tilde{\rho}_1(\omega)-1\\
     2\tilde{\rho}_{2}(\omega)&=\tilde{a}_{\frac{1}{2}}(\omega)+2\tilde{K}^{\{2\}}(\omega)\tilde{\rho}_1(\omega)-2\tilde{K}^{\{1\}}(\omega)\tilde{\rho}_2(\omega)-1-2\sum_{i=1}^3\cos(\omega\vartheta_i).
\end{align}
The solution of the above coupled algebraic equations is
\begin{align}
    \tilde\rho_1(\omega)&=\frac{e^{-\frac{| \omega | }{2}}-e^{| \omega | }-2 e^{\frac{| \omega | }{2}} \left(\sum_{i=1}^3\cos \left(\omega  \vartheta_i\right)\right)+4 \cosh \left(\frac{| \omega | }{2}\right) \left(\cos (d \omega )+\cos (b
   \omega ) N_q\right)}{2 (1+2 \cosh (| \omega | ))}\\
   \tilde\rho_2(\omega)&=\frac{e^{-\frac{| \omega | }{2}}-e^{| \omega | }+2 \left(\cos (d \omega )-\left(1+e^{| \omega | }\right) \left(\sum_{i=1}^3\cos \left(\omega  \vartheta_i\right)\right)+\cos (b \omega ) N_q\right)}{2 (1+2 \cosh (|
   \omega | ))}
\end{align}

As a result, the total number of rank‐1 roots becomes
\begin{equation}
M_{1} = \int \rho_{1}(\lambda) d\lambda = \frac{1}{3} \left(2 N_q-1\right),
\end{equation}
while the total number of rank‐2 roots is
\begin{equation}
M_{2} = \int \rho_{2}(\mu) d\mu = \frac{1}{3} \left(N_q-5\right).
\end{equation}
Thus, this excitation belongs to a 10-dimensional multiple with Dynkin labels (0,3) and $\mathbb{Z}_3$ triality 0. The change in the density of the rank-1 equation due to the presence of these three holes in the rank-2 equation is
\begin{equation}
    \Delta \tilde{\rho}(\vartheta_1,\vartheta_2,\vartheta_3)=-\frac{e^{\frac{| \omega | }{2}} \sum_{i=1}^3\cos (\omega \vartheta_i  )}{2
   \cosh (\omega )+1},
\end{equation}
such that the energy of this excitation is
\begin{equation}
    E_{\vartheta_1,\vartheta_2,\vartheta_3}=m\sum_{i=1}^3\cosh\left(\frac{2\pi}{3}\vartheta_i\right),
\end{equation}
which attains the minimum value of $3m$ as the rapidities of the holes approach infinity. 

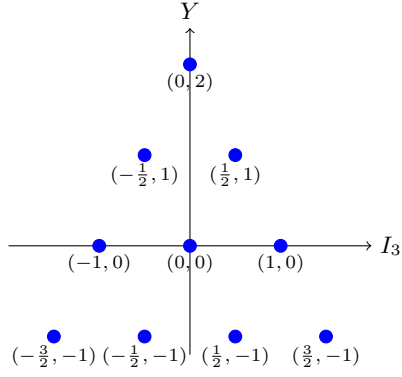
\begin{figure}[H]
    \centering
\begin{tikzpicture}[scale=1.2]
  \draw[->] (-2,0) -- (2,0) node[right] {$I_3$};
  \draw[->] (0,-1.2) -- (0,2.4) node[above] {$Y$};
  \foreach \x/\y/\label in {
    0/2/{$(0,2)$},
    0.5/1/{$(\frac{1}{2},1)$}, -0.5/1/{$(-\frac{1}{2},1)$},
    1/0/{$(1,0)$}, 0/0/{$(0,0)$}, -1/0/{$(-1,0)$},
    1.5/-1/{$(\frac{3}{2},-1)$}, 0.5/-1/{$(\frac{1}{2},-1)$}, -0.5/-1/{$(-\frac{1}{2},-1)$}, -1.5/-1/{$(-\frac{3}{2},-1)$}}
    {
      \filldraw[blue] (\x,\y) circle (2pt);
      \node[anchor=north] at (\x,\y) {\scriptsize \label};
    }
\end{tikzpicture}
    \caption{Weight diagram of the $SU(3)$ decuplet $(0,3)$ of anti-baryons plotted in the hypercharge $Y$ versus isospin $I_3$ plane.}
    \label{fig:baryob-antidecuplet1}
\end{figure}

These excitations carry Dynkin labels $(p,q)=(0,3)$, which is 10-dimensional, and the highest weight state carries $Y_{\rm max}=2$ and $I_3=0$. The highest weight state and the nine descendant states are plotted in the $Y$–$I_3$ plane in Fig.~\ref{fig:baryob-antidecuplet1}.

\subsubsection{DOS and the Kondo temperature $T_K$}

In the ground state, the density of the rank-1 root in Fourier space is given by
\begin{equation}
    \widetilde\rho_1(\omega)
=\frac{e^{-|\omega|/2}\left[
2\left(e^{|\omega|}+1\right)N_q\cos(b|\omega|)
+2\left(e^{|\omega|}+1\right)\cos(d|\omega|)
-e^{3|\omega|/2}+1
\right]}{2\left(2\cosh|\omega|+1\right)}.
\end{equation}

The contribution of the bulk degrees of freedom to the density of the rank-1 roots is
\begin{equation}
    \widetilde\rho_1^{\rm bulk}(\omega)
=\frac{2N_q\cosh(\omega/2)\cos(b\omega)}{2\cosh(\omega)+1},
\end{equation}
which can be written in the Fourier space as
$$
\rho_1(\lambda)
=\frac{N_q}{2\sqrt3}\left[
\frac{1}{2\cosh\!\left(\frac{2\pi}{3}(b+\lambda)\right)-1}
+\frac{1}{2\cosh\!\left(\frac{2\pi}{3}(b-\lambda)\right)-1}
\right].
$$

From the energy of a single fundamental hole, one obtains
$$
E'(\lambda)
=\frac{2\pi}{3}m\sinh\!\left(\frac{2\pi\lambda}{3}\right).
$$
Hence the bulk DOS is
\begin{equation}
    \mathrm{DOS}_{\rm bulk}=\frac{\rho_1(\lambda)}{E'(\lambda)}
=N^q
\frac{
\left[\frac{1}{2\cosh\!\left(\frac{2\pi(b+\lambda)}{3}\right)-1}
+\frac{1}{2\cosh\!\left(\frac{2\pi(b-\lambda)}{3}\right)-1}\right]}
{\frac{2\pi}{3}m\sin\!\left(\frac{2\pi\lambda}{3}\right)}.
\end{equation}
In the limit $b\to\infty$, using $e^{-2\pi b/3}=\frac{m}{2\sqrt{3}D}$, this becomes
$$
\mathrm{DOS}_{\rm bulk}(E)
=\frac{NE}{4\pi D\sqrt{E^2-m^2}}=\frac{L}{4\pi}\frac{E}{\sqrt{E^2-m^2}}.
$$

The impurity DOS is
\begin{equation}
    \mathrm{DOS}_{\rm imp}(E)
=\frac{4\sqrt3E\cosh\!\left(\frac{2\pi d}{3}\right)
-2\sqrt3m}
{\pi m
\sqrt{12E^2-12m^2}
\left[2\cosh\!\left(\frac{2\pi d}{3}-\cosh^{-1}\!\frac{E}{m}\right)-1\right]
\left[2\cosh\!\left(\frac{2\pi d}{3}+\cosh^{-1}\!\frac{E}{m}\right)-1\right]}.
\end{equation}
In the limit $d\gg1$, one obtains
\begin{equation}
    \mathrm{DOS}_{\rm imp}(E)
\approx
\frac{2\sqrt3me^{2\pi d/3}E}
{2\pi m\sqrt{E^2-m^2}\left(e^{4\pi d/3}m^2+4E^2\right)}
=\frac{\sqrt3e^{2\pi d/3}E}{\pi\sqrt{E^2-m^2}\left(e^{4\pi d/3}m^2+4E^2\right)}.
\end{equation}

Defining
\begin{equation}
    T_K=\frac{e^{2\pi d/3}m}{2},
\end{equation}
so that $e^{4\pi d/3}m^2=4T_K^2$, we obtain
\begin{equation}
    R(E)
=\frac{\sqrt3}{2\pi}\frac{T_K}{E^2+T_K^2}.
\end{equation}

\subsection{The YSR phase}\label{ysr-detailed}
Notice that the rank-1 Bethe Ansatz equations change when $\delta>\frac{1}{2}$ and takes the from
\begin{align}
\prod_{\upsilon=\pm} \left[
\left( \frac{\lambda_\alpha + \upsilon b - \frac{i}{2}}{\lambda_\alpha + \upsilon b + \frac{i}{2}} \right)^{N_q}
\frac{\lambda_\alpha +  i(\delta - \frac{1}{2})}{\lambda_\alpha +  i(\delta + \frac{1}{2})}\frac{\lambda_\alpha - i(\delta + \frac{1}{2})}{\lambda_\alpha + i(\delta - \frac{1}{2})}
\prod_{k=1}^{M_2}
\frac{\lambda_\alpha + \upsilon \mu_k - \frac{i}{2}}{\lambda_\alpha + \upsilon \mu_k + \frac{i}{2}}
\right]
&=
\prod_{\substack{\beta=1 \\ \beta \ne \alpha}}^{M_1}
\prod_{\upsilon=\pm}
\frac{\lambda_\alpha + \upsilon \lambda_\beta - i}{\lambda_\alpha + \upsilon \lambda_\beta + i}
\quad\quad \alpha = 1,\dots, M_1,
\label{rank1BAE-BM}
\end{align}
whereas the rank-2 equation remains unchanged \textit{i.e.}
\begin{equation}
\begin{aligned}
\prod_{\alpha=1}^{M_1} \prod_{\upsilon=\pm}
\frac{\mu_k + \upsilon \lambda_\alpha - \frac{i}{2}}{\mu_k + \upsilon \lambda_\alpha + \frac{i}{2}}
&=
\prod_{\substack{\ell=1 \\ \ell \ne k}}^{M_2}
\prod_{\upsilon=\pm}
\frac{\mu_k + \upsilon \mu_\ell - i}{\mu_k + \upsilon \mu_\ell + i}
\quad\quad k = 1,\dots, M_2
\end{aligned}
\label{rank2-BAE-bm}
\end{equation}

Now, the logarithm of the two Bethe ranks of Bethe Ansatz equations read
\begin{align}
    &\sum_{\upsilon=\pm}\left(N_q \tan^{-1}(2 (\lambda_\alpha+ \upsilon b))+\tan ^{-1}\left(\frac{2 \lambda_\alpha }{1+2 \delta 
   \upsilon }\right) +\sum_{k=1}^{M_2}\tan^{-1}(2(\lambda_\alpha+\upsilon\mu_k))\right)+\tan^{-1}(2\lambda_\alpha)\nonumber\\
   &\quad\quad\quad=\pi I_\alpha +\sum_{\upsilon=\pm}\sum_{\beta}\tan^{-1}(\lambda_\alpha+\upsilon\lambda_\beta)
   \label{rank1EQN--1}
\end{align}
where one should notice that when $\upsilon=-1$, the $\tan^{-1}$ term involving the impurity parameter $\delta$ is negative for $\delta>\frac{1}{2}$.
and

\begin{equation}
    \tan^{-1}(2\mu_k)+\sum_{\upsilon=\pm} \sum_{\alpha=1}^{M_1}\tan^{-1}(2(\mu_k+\upsilon \lambda_\alpha))=\sum_{\upsilon=\pm}\sum_{\ell=1}^{M_2}\tan^{-1}(\mu_k+\upsilon\mu_{\ell})+ \pi I_{k}
    \label{rank2EQN--2}
\end{equation}
where $I_\alpha$ and $I_k$ are integers that represent the color quantum numbers. 

Differentiating with respect to $\lambda_\alpha$ (and similarly $\mu_k$) turns each
$\tan^{-1}(x)$ into the Lorentzian $\frac{d}{dx}\tan^{-1}(x)=\frac{1}{1+x^2}$,
and the two equations become
\begin{align}
2 \rho_1(\lambda)&= g(\lambda)+\sum_{\upsilon=\pm}\left(\int K^{\{2\}} (\lambda+\upsilon\mu)\rho_2(\mu)\mathrm{d}\mu-\int K^{\{1\}}(\lambda+\upsilon\lambda')\rho_1(\lambda')\mathrm{d}\lambda'\right)-\delta(\lambda),\\
2\rho_2(\mu)&= a_{\frac{1}{2}}(\mu)+\sum_{\upsilon=\pm}\left(\int K^{\{2\}}(\mu+\upsilon \lambda)\rho_1(\lambda)\mathrm{d}\lambda - \int K^{\{1\}}(\mu+\upsilon \mu')\mathrm{d}\mu'\right)-\delta(\mu)
\end{align}
where the two delta functions are added to remove the two roots $\lambda_\alpha=0$ and $\mu_k=0$, which leads to un-normalizable solutions. Moreover, here
\begin{align}
&g(\lambda)=a_{\frac{1}{2}}(\lambda)+\left(N_q \sum_{\upsilon=\pm}a_{\frac{1}{2}}(\lambda+\upsilon b) -a_{\delta-\frac{1}{2}}(\lambda)+a_{\delta+\frac{1}{2}}(\lambda)\right)
\end{align}

Now, we can write these two equations in Fourier space as
\begin{align}
    2\tilde{\rho}_{1}(\omega)&=\tilde g (\omega)+ 2\tilde{K}^{\{2\}}(\omega)\tilde{\rho}_2(\omega)-2\tilde{K}^{\{1\}}(\omega)\tilde{\rho}_1(\omega)-1\\
     2\tilde{\rho}_{2}(\omega)&=\tilde{a}_{\frac{1}{2}}(\omega)+2\tilde{K}^{\{2\}}(\omega)\tilde{\rho}_1(\omega)-2\tilde{K}^{\{1\}}(\omega)\tilde{\rho}_2(\omega)-1
\end{align}
Solving this, we get
\begin{align}
    \tilde\rho_1(\omega)&=\frac{4 N_q \cosh \left(\frac{| \omega | }{2}\right) \cos (b \omega )+\left(e^{2 | \omega | }-1\right) \left(-e^{-\frac{1}{2} (2 \delta
   +1) | \omega | }\right)+e^{-\frac{| \omega | }{2}}-e^{| \omega | }}{2 (2 \cosh (| \omega | )+1)}\\
   \tilde\rho_2(\omega)&=\frac{e^{-\frac{| \omega | }{2}}-e^{| \omega | }-e^{-\delta 
   | \omega | } \left(-1+e^{| \omega | }\right)+2 \cos (b
   \omega ) N_q}{2 (1+2 \cosh (| \omega | ))}
\end{align}
such that $M_1=\int \mathrm{d}\lambda \tilde{\rho}_1(\lambda)=\frac{2N_q}{3}$ and $M_2=\int \mathrm{d}\mu \tilde{\rho}_1(\mu)=\frac{N_q}{3}$. Hence, the Dynkin labels of this state is $(1,0)$. This single fundamental quark excitation is due to the fact that the impurity quark is unscreened in this state. However, when $\frac{1}{2}<\delta<2$, it is possible to construct a low energy eigenstate where impurity is screened by a single particle bound mode. 

The construction of this state is different when $\frac{1}{2}<\delta<1$ and when $1<\delta<2$. In the former case, there exists a unique solution of Bethe Ansatz equations of rank-1 of the form
\begin{equation}
    \lambda_\delta=\pm i\left(\delta-\frac12\right).
\end{equation}

Adding this solution, we obtain (for parameter range $1/2<\delta<1$), both rank-1 and rank-2 equation changes as follows:

\begin{align}
&\prod_{\upsilon=\pm} \left[
\left( \frac{\lambda_\alpha + \upsilon b - \frac{i}{2}}{\lambda_\alpha + \upsilon b + \frac{i}{2}} \right)^{N_q}
\frac{\lambda_\alpha +  i(\delta - \frac{1}{2})}{\lambda_\alpha +  i(\delta + \frac{1}{2})}\frac{\lambda_\alpha +i(\frac{3}{2}-\delta)}{\lambda_\alpha - i(\frac{3}{2}-\delta)}
\prod_{k=1}^{M_2}
\frac{\lambda_\alpha + \upsilon \mu_k - \frac{i}{2}}{\lambda_\alpha + \upsilon \mu_k + \frac{i}{2}}
\right]=\prod_{\substack{\beta=1 \\ \beta \ne \alpha}}^{M_1}
\prod_{\upsilon=\pm}
\frac{\lambda_\alpha + \upsilon \lambda_\beta - i}{\lambda_\alpha + \upsilon \lambda_\beta + i}
\quad\quad \alpha = 1,\dots, M_1,
\label{rank1BAE-BM-BS}
\end{align}
and
\begin{equation}
\begin{aligned}
\frac{\left(\mu_{k} - i \delta\right)}{\left(\mu_{k} + i \delta\right)}
\frac{\left(\mu_{k} - i\left(1 - \delta\right)\right)}{\left(\mu_{k} + i\left(1 - \delta\right)\right)}\prod_{\alpha=1}^{M_1} \prod_{\upsilon=\pm}
\frac{\mu_k + \upsilon \lambda_\alpha - \frac{i}{2}}{\mu_k + \upsilon \lambda_\alpha + \frac{i}{2}}
&=
\prod_{\substack{\ell=1 \\ \ell \ne k}}^{M_2}
\prod_{\upsilon=\pm}
\frac{\mu_k + \upsilon \mu_\ell - i}{\mu_k + \upsilon \mu_\ell + i}
\quad\quad k = 1,\dots, M_2
\end{aligned}
\label{rank2-BAE-bm-bs}
\end{equation}

The two Bethe Ansatz equations can be written in Fourier space as

\begin{align}
    2\tilde{\rho}_{1}(\omega)&=\tilde{a}_{\frac{1}{2}}(\omega)(1+2N_q\cos(\omega b))+ 2\tilde{K}^{\{2\}}(\omega)\tilde{\rho}_2(\omega)-2\tilde{K}^{\{1\}}(\omega)\tilde{\rho}_1(\omega)-1-e^{\frac{| \omega | }{2}-\delta  | \omega | }-e^{\delta  |
   \omega | -\frac{3 | \omega | }{2}}\\
     2\tilde{\rho}_{2}(\omega)&=\tilde{a}_{\frac{1}{2}}(\omega)+2\tilde{K}^{\{2\}}(\omega)\tilde{\rho}_1(\omega)-2\tilde{K}^{\{1\}}(\omega)\tilde{\rho}_2(\omega)-1+e^{-((1-\delta ) | \omega | )}+e^{-\delta  | \omega | }
\end{align}
and the solution of coupled algebraic equations leads to the solution of the root density of the form
\begin{align}
\tilde \rho_1(\omega)&=\frac{4 N_q \cosh \left(\frac{| \omega | }{2}\right) \cos (b \omega )-2 \cosh \left(\frac{1}{2} (3-2 \delta ) | \omega |
   \right)+e^{-\frac{| \omega | }{2}}-e^{| \omega | }}{2 (2 \cosh (| \omega | )+1)}\label{sol-rank1-bm-1st}\\
\tilde \rho_2(\omega)&=\frac{2 \cosh (\delta  | \omega | )+e^{-\frac{| \omega | }{2}}-e^{| \omega | }+2 N_q \cos (b \omega )}{2 (2 \cosh (| \omega | )+1)},\label{sol-rank2-bm-1st}
\end{align}
such that there are
\begin{equation}
    M_1=1+\int \rho_1(\lambda)\mathrm{d}\lambda=\tilde\rho_1(0)=\frac{2}{3} \left(N_q+1\right)
\end{equation} 
number of roots of the first rank of equations, and there are
\begin{equation}
    M_2=\int \rho_2(\mu)\mathrm{d}\mu=\tilde\rho_2(0)=\frac{1}{3} \left(N_q+1\right)
\end{equation}

The Dynkin labels of the states are $(p,q)=(0,0)$,
which shows that in this state the impurity is screened by a single particle bound mode. The change in the density of rank-1 roots due to the presence of this 1 purely imaginary root in the rank-1 equation is
\begin{equation}
    \Delta\rho_1(\omega)=-\frac{e^{\frac{1}{2} (2 \delta -1) | \omega | } \left(e^{(1-2 \delta ) | \omega | }+1\right)}{2 \left(e^{| \omega | }+e^{2 | \omega |
   }+1\right)}
   \label{change1st}
\end{equation}

Before we compute the energy of the bound mode, we now turn to the construction of this state for the parametric range $1<\delta<2$. In this regime, the rank-2 equation given by Eq.~\eqref{rank2-BAE-bm-bs} becomes
\begin{align}
\frac{\left(\mu_{k} - i \delta\right)}{\left(\mu_{k} + i \delta\right)}
\frac{\left(\mu_{k} + i\left(\delta-1\right)\right)}{\left(\mu_{k} -i\left( \delta-1\right)\right)}\prod_{\alpha=1}^{M_1} \prod_{\upsilon=\pm}
\frac{\mu_k + \upsilon \lambda_\alpha - \frac{i}{2}}{\mu_k + \upsilon \lambda_\alpha + \frac{i}{2}}
&=
\prod_{\substack{\ell=1 \\ \ell \ne k}}^{M_2}
\prod_{\upsilon=\pm}
\frac{\mu_k + \upsilon \mu_\ell - i}{\mu_k + \upsilon \mu_\ell + i}
\quad\quad k = 1,\dots, M_2,
\end{align}
such that there exists a boundary string of the form
\begin{equation}
    \mu+k=\pm i(\delta-1)
\end{equation}
in the rank-2 equation. Adding this solution, the rank-2 equation becomes
\begin{align}
\frac{\left(\mu_{k} + i\left(\delta-1\right)\right)}{\left(\mu_{k} -i\left( \delta-1\right)\right)}\prod_{\alpha=1}^{M_1} \prod_{\upsilon=\pm}
\frac{\mu_k + \upsilon \lambda_\alpha - \frac{i}{2}}{\mu_k + \upsilon \lambda_\alpha + \frac{i}{2}}
&=
\frac{\mu_k - i (2 - \delta)}{\mu_k + i (2 - \delta)}\prod_{\ell=1}^{M_2}
  \frac{(v_k \pm v_\ell)-i}{(v_k \pm v_\ell)+i}\prod_{\substack{\ell=1 \\ \ell \ne k}}^{M_2}
\prod_{\upsilon=\pm}
\frac{\mu_k + \upsilon \mu_\ell - i}{\mu_k + \upsilon \mu_\ell + i}.
\label{rank-2-eqn-bm-2bs}
\end{align}

This then changes the rank-1 equation to
\begin{equation}
    \prod_{\upsilon=\pm} \left[
\left( \frac{\lambda_\alpha + \upsilon b - \frac{i}{2}}{\lambda_\alpha + \upsilon b + \frac{i}{2}} \right)^{N_q}
\prod_{k=1}^{M_2}
\frac{\lambda_\alpha + \upsilon \mu_k - \frac{i}{2}}{\lambda_\alpha + \upsilon \mu_k + \frac{i}{2}}
\right]=
\prod_{\substack{\beta=1 \\ \beta \ne \alpha}}^{M_1}
\prod_{\upsilon=\pm}
\frac{\lambda_\alpha + \upsilon \lambda_\beta - i}{\lambda_\alpha + \upsilon \lambda_\beta + i}
\quad\quad \alpha = 1,\dots, M_1.
\label{rank-1-eqn-bm-2bs}
\end{equation}

The coupled Bethe Ansatz equations can thus be written in the Fourier space as
\begin{align}
    2\tilde{\rho}_{1}(\omega)&=\tilde{a}_{\frac{1}{2}}(\omega)(1+2N_q\cos(\omega b))+ 2\tilde{K}^{\{2\}}(\omega)\tilde{\rho}_2(\omega)-2\tilde{K}^{\{1\}}(\omega)\tilde{\rho}_1(\omega)-1\\
     2\tilde{\rho}_{2}(\omega)&=\tilde{a}_{\frac{1}{2}}(\omega)+2\tilde{K}^{\{2\}}(\omega)\tilde{\rho}_1(\omega)-2\tilde{K}^{\{1\}}(\omega)\tilde{\rho}_2(\omega)-1-e^{-((2-\delta ) | \omega | )}-e^{-((\delta -1) | \omega | )}.
\end{align}

The solution of these coupled equations is of the form
\begin{align}
    \tilde\rho_1(\omega)&=\frac{4 N_q \cosh \left(\frac{| \omega | }{2}\right) \cos (b \omega )-2 \cosh \left(\frac{1}{2} (3-2 \delta ) | \omega |
   \right)+e^{-\frac{| \omega | }{2}}-e^{| \omega | }}{2 (2 \cosh (| \omega | )+1)}\label{sol-rank1-bm-2st}\\
   \tilde\rho_2(\omega)&=\frac{-2 \cosh ((\delta -2) | \omega | )-2 \cosh ((\delta -1) | \omega | )+e^{-\frac{| \omega | }{2}}-e^{| \omega | }+2 N_q \cos (b
   \omega )}{2 (2 \cosh (| \omega | )+1)}.\label{sol-rank2-bm-2st}
\end{align}

The total number of rank-1 roots (including the one purely imaginary boundary string root) in this state is $M_1=1+\int\rho_1(\lambda)\mathrm{d}\lambda=\frac{2}{3}(N_q+1)$ and total number of rank-2 roots (including the one purely imaginary boundary string root)  is  $M_1=1+\int\rho_2(\mu)\mathrm{d}\mu=\frac{1}{3}(N_q+1)$. The Dynkin labels of this state are (0,0) which suggests that this is a state with $\mathbb{Z}_3$ triality 0
state where impurity is screened by a localized bound mode. Notice that while the solutions to the Bethe Ansatz equations containing the boundary bound modes Eq.~\eqref{sol-rank1-bm-2st} and Eq.~\eqref{sol-rank2-bm-2st} are different than those obtained in regime $\frac{1}{2}<\delta<1$ where the solutions of the Bethe Ansatz equations in Eq.~\eqref{sol-rank1-bm-1st} and Eq.~\eqref{sol-rank2-bm-1st}, the change in the density of the rank-1 roots due to the presence of these two boundary bound strings solution is exactly the same as the one given by Eq.~\eqref{change1st} i.e. the change in root density of rank-1 is
\begin{equation}
    \Delta\rho_1(\omega)=-\frac{e^{\frac{1}{2} (2 \delta -1) | \omega | } \left(e^{(1-2 \delta ) | \omega | }+1\right)}{2 \left(e^{| \omega | }+e^{2 | \omega |
   }+1\right)}.
\end{equation}

The energy of the boundary bound mode is then obtained by adding the bare energy of the rank-1 boundary string and the energy due to the change in the root density of the rank-1 solution using Eq.~\eqref{engeqn}. We obtain
\begin{align}
    E_\delta&=-D \tan ^{-1}\left(\frac{b}{b^2+(\delta -1) \delta }\right)+\frac{D}{2\pi}\int  \left(-\frac{e^{-\frac{1}{2} (1-2 \delta ) | \omega | }
   \left(e^{(1-2 \delta ) | \omega | }+1\right)}{2 \left(e^{|
   \omega | }+e^{2 | \omega | }+1\right)}\right)\left(\frac{2i\pi}{\omega}e^{-\frac12|\omega|-ib\omega}\right)\mathrm{d}\omega\nonumber\\
   &=-D \tan^{-1}\left(
\frac{2 \sin\left(\frac{\pi}{6}(4\delta+1)\right) - \cosh\left(\frac{2\pi b}{3}\right)}
     {\sqrt{3} \sinh\left(\frac{2\pi b}{3}\right)}
\right)
\end{align}

Taking the scaling limit $D\to \infty$ and $b\to\infty$, we find that the energy of this single particle bound mode is
\begin{equation}
    E_\delta=-m\sin\left(\frac{\pi}{6}(1+4\delta)\right).
\end{equation}

As shown in the main text, the energy of this mode is negative when $\frac12<\delta<\frac54$ and it is positive when $\frac54<\delta<2$. Thus, the $\mathbb{Z}_3$ triality 0 state with Dynkin labels (0,0) where the impurity is screened by the bound mode is the ground state in the regime $\frac12<\delta<\frac54$ whereas in the regime $\frac54<\delta<2$, the $\mathbb{Z}_3$ triality 1 state with Dynkin labels (1,0) where the impurity quark is unscreened is the ground state. There exists a mid-gap state with energy $E_\delta$ in the $\frac12<\delta<\frac54$ where the impurity quark is unscreened, and this state is a triality 1 state with Dynkin labels(1,0). Likewise, there exists a mid-gap state with energy $E_\delta$ in the regime $\frac54<\delta<2$, which is a state with Dynkin labels(0,0) and triality 0 where the impurity quark is screened by a single particle bound mode. 

As discussed in the main text, starting from the $\mathbb{Z}_3$ triality 1 state with unscreened impurity quark, one can add a single hole at position $\vartheta$ to construct a state with Dynkin labels (1,1) and triality 0 where the unscreened quark and an itinerant bulk antiquark forms an 8-dimensional meson octet state with energy $E_{\vartheta}=m\cosh\left(\frac{2\pi}{3}\vartheta\right)$ which attains its minimum value of $m$ when $\vartheta\to \infty$. Likewise, one can also construct a state with Dynkin labels $(3,0)$ by adding two holes at positions $\theta_1$ and $\theta_2$ at the rank-1 equation such that the unscreened quark and the two itinerant bulk quarks form baryon decuplet. This state has energy $E_{\theta_1,\theta_2}=m\left(\cosh\left(\frac{2\pi}{3}\theta_1\right)+\cosh\left(\frac{2\pi}{3}\theta_2\right)\right)$ which attains the minimum value of $2m$ when $\theta_i\to \infty ~\forall i=\{1,2\}.$

All the excitations constructed in Sec.\ref{fund-ext-app} are valid excitations in the bulk in this regime also. And the above-constructed boundary excitations are not all possible boundary expectations. For example, one unique triality 2 boundary excitation is possible in the parametric regime $1<\delta<3/2$ by adding only the boundary string in the rank-1 Bethe equations such that the Bethe Ansatz equations in Fourier space become
\begin{align}
    2\tilde{\rho}_{1}(\omega)&=\tilde{a}_{\frac{1}{2}}(\omega)(1+2N_q\cos(\omega b))+ 2\tilde{K}^{\{2\}}(\omega)\tilde{\rho}_2(\omega)-2\tilde{K}^{\{1\}}(\omega)\tilde{\rho}_1(\omega)-1-e^{\frac{| \omega | }{2}-\delta  | \omega | }-e^{\delta  |
   \omega | -\frac{3 | \omega | }{2}}\\
     2\tilde{\rho}_{2}(\omega)&=\tilde{a}_{\frac{1}{2}}(\omega)+2\tilde{K}^{\{2\}}(\omega)\tilde{\rho}_1(\omega)-2\tilde{K}^{\{1\}}(\omega)\tilde{\rho}_2(\omega)-1-e^{-((\delta-1 ) | \omega | )}+e^{-\delta  | \omega | }.
\end{align}

The solution of these coupled equations is of the form
\begin{align}
    \tilde\rho_1(\omega)&=\frac{4 N_b \cosh \left(\frac{\omega }{2}\right) \cos (b \omega )+e^{-\frac{| \omega | }{2}}-e^{| \omega | }+e^{\delta  | \omega |
   -\frac{3 | \omega | }{2}} \left(-2 e^{(3-2 \delta ) | \omega | }+e^{| \omega | }+1\right)}{2 (2 \cosh (\omega )+1)}\label{eqn1-antq}\\
   \tilde\rho_2(\omega)&=\frac{2 \cos (b \omega ) N_b+e^{-\frac{| \omega | }{2}}-e^{| \omega | }-e^{-((-2+\delta ) | \omega | )}+e^{-\delta  | \omega | }-2
   \cosh ((-1+\delta ) \omega )}{2 (1+2 \cosh (\omega ))},\label{eqn2-antq}
\end{align}
such that the total number of rank-1 roots including the one purely imaginary boundary string solution is $M_1=1+\int\rho_1(\lambda)\mathrm{d}\lambda=\frac{1}{3} \left(2 N_q+1\right)$ and the number of rant-2 roots is $M_2=\int\rho_2(\mu)\mathrm{d}\mu=\frac{1}{3} \left(N_q-1\right)$. The Dynkin labels of this state is $(p,q)=(0,1)$. This shows that this is a triality 2 state with an unscreened antiquark at the boundary.

The change in the density of the rank 1 roots due to the presence of this boundary string solution is
\begin{equation}
    \Delta\tilde{\rho}_1(\omega)=-\frac{e^{\delta  | \omega | -\frac{3 | \omega | }{2}} \left(e^{(1-2 \delta ) | \omega | }+e^{(3-2 \delta ) | \omega | }+e^{| \omega |
   }+1\right)}{2 (2 \cosh (\omega )+1)}
\end{equation}

The energy of this excitation is
\begin{align}
    E_\delta&=-D \tan ^{-1}\left(\frac{b}{b^2+(\delta -1) \delta }\right)+\frac{D}{2\pi}\int  \left(-\frac{e^{\delta  | \omega | -\frac{3 | \omega | }{2}} \left(e^{(1-2 \delta ) | \omega | }+e^{(3-2 \delta ) | \omega | }+e^{| \omega |
   }+1\right)}{2 (2 \cosh (\omega )+1)}\right)\left(\frac{2i\pi}{\omega}e^{-\frac12|\omega|-ib\omega}\right)\mathrm{d}\omega\nonumber\\
   &=\frac{1}{2} D i \log \left(\frac{\cos \left(\frac{2 \pi  \delta }{3}\right)+\cos \left(\frac{1}{3} (\pi +2 i \pi  b)\right)}{\cos
   \left(\frac{2 \pi  \delta }{3}\right)+\sin \left(\frac{1}{6} \pi  (1+4 i b)\right)}\right).
\end{align}
Taking the double scaling limit $D\to\infty$ and $b\to\infty$, we get the energy of this antiquark excitation is
\begin{equation}
    E_\delta=-m\cos\left(\frac{2\pi}{3}\delta\right).
\end{equation}

Likewise, in the parametric regime $\delta>\frac{3}{2}$, adding only the rank 1 boundary string leads to the Bethe Ansatz equations for the root densities of the form
\begin{align}
    2\tilde{\rho}_{1}(\omega)&=\tilde{a}_{\frac{1}{2}}(\omega)(1+2N_q\cos(\omega b))+ 2\tilde{K}^{\{2\}}(\omega)\tilde{\rho}_2(\omega)-2\tilde{K}^{\{1\}}(\omega)\tilde{\rho}_1(\omega)-1-e^{\frac{| \omega | }{2}-\delta  | \omega | }+e^{-\delta  |
   \omega | +\frac{3 | \omega | }{2}}\\
     2\tilde{\rho}_{2}(\omega)&=\tilde{a}_{\frac{1}{2}}(\omega)+2\tilde{K}^{\{2\}}(\omega)\tilde{\rho}_1(\omega)-2\tilde{K}^{\{1\}}(\omega)\tilde{\rho}_2(\omega)-1-e^{-((\delta-1 ) | \omega | )}+e^{-\delta  | \omega | }.
\end{align}
This leads to a solution with negative Dynkin label. However, adding a hole in rank 1 equations leads to the solution of the from
\begin{align}
    \tilde{\rho}_1(\omega))&=\frac{\left(e^{| \omega | }-1\right) e^{\frac{1}{2} (3-2 \delta ) | \omega | }-2 \left(e^{| \omega | }+1\right) \cos (\theta  \omega
   )+e^{-\frac{| \omega | }{2}}-e^{| \omega | }+4 N_q \cosh \left(\frac{\omega }{2}\right) \cos (b \omega )}{2 (2 \cosh (\omega )+1)}\\
   \tilde{\rho}_2(\omega))&\frac{\left(e^{| \omega | }-1\right) \left(-e^{-\delta  | \omega | }\right)-2 e^{\frac{| \omega | }{2}} \cos (\theta  \omega
   )+e^{-\frac{| \omega | }{2}}-e^{| \omega | }+2 N_q \cos (b \omega )}{2 (2 \cosh (\omega )+1)}.
\end{align}

The total number of rank-1 roots including the one purely imaginary boundary string solution is $M_1=1+\int\rho_1(\lambda)\mathrm{d}\lambda=\frac{1}{3} \left(2 N_q+1\right)$ and the number of rant-2 roots is $M_2=\int\rho_2(\mu)\mathrm{d}\mu=\frac{1}{6} \left(2 N_q-2\right)$. Thus, the Dynkin label of this state is $(p,q)=(0,1)$. Hence, this constitutes an additional triality 2 excitation, transforming in the antifundamental ($\bar{\mathbf{3}}$) representation of $\mathrm{SU}(3)$. The change in the density of rank 1 roots due to the presence of the boundary string solution and the hole in rank 1 equation is
\begin{equation}
    \Delta\tilde\rho_1(\omega)=\frac{1}{2} \left(e^{| \omega | }-1\right) e^{\frac{1}{2} (1-2 \delta ) | \omega | }+\frac{\left(e^{| \omega | }+1\right) \cos (\theta 
   | \omega | )}{2 \cosh (| \omega | )+1},
\end{equation}
such that the total energy of this excitation is
\begin{equation}
    E_\theta=-\tan ^{-1}\left(\frac{b}{b^2+(\delta -1) \delta }\right)+\frac{D}{2\pi}\int \left(\frac{1}{2} \left(e^{| \omega | }-1\right) e^{\frac{1}{2} (1-2 \delta ) | \omega | }+\frac{\left(e^{| \omega | }+1\right) \cos (\theta 
   | \omega | )}{2 \cosh (| \omega | )+1} \right)\left(\frac{2i\pi}{\omega}e^{-\frac12|\omega|-ib\omega} \right) \mathrm{d}\omega.
\end{equation}
The first term is the bare energy which exactly cancels the energy contribution from the back flow of the roots due to the presence of the boundary string solutions. Thus, the total energy comes solely from the energy of the hole which in the scaling limit becomes
\begin{equation}
    E=m\cosh\left(\frac{2\pi}{3}\theta\right).
\end{equation}

\subsection{The Unscreened regime}
As the parameter $\delta$ further increases and takes a value in the range $\delta>2$, the bulk superconducting scale overwhelms the boundary Kondo scale and hence the impurity is no longer screened. Or in other words, the beta function of the boundary coupling $J$ reverses its sign and becomes positive, and hence flows to weak coupling where the impurity quark is unscreened. 

In this regime, when the two Bethe Ansatz equations for all real roots of both rank equations are
\begin{align}
2\tilde{\rho}_{1}(\omega) &= \tilde a_{\frac12}(\omega)(1+\cos(\omega b))+\tilde a_{\delta-\frac{1}{2}}(\omega)-\tilde a_{\delta+\frac12}(\omega)
+ 2\tilde K^{\{2\}}(\omega)\tilde{\rho}_{2}(\omega)
- 2\tilde K^{\{1\}}(\omega)\tilde{\rho}_{1}(\omega)
- 1,\\
2\tilde{\rho}_{2}(\omega) &= \tilde a_{\frac{1}{2}}(\omega)
+ 2\tilde K^{\{2\}}(\omega)\tilde{\rho}_{1}(\omega)
- 2\tilde K^{\{1\}}(\omega)\tilde{\rho}_{2}(\omega)
- 1.
\end{align}
Solving these equations gives
\begin{align}
\tilde{\rho}_{1}(\omega)
&=
\frac{4N_{q}\cosh\left(\frac{|\omega|}{2}\right)\cos\left(b\omega\right)
+\left(e^{2|\omega|}-1\right)\left(-e^{-\frac{1}{2}(2\delta+1)|\omega|}\right)
+e^{-\frac{|\omega|}{2}}-e^{|\omega|}}
{2\left(2\cosh\left(|\omega|\right)+1\right)},\\
\tilde{\rho}_{2}(\omega)
&=
\frac{e^{-\frac{|\omega|}{2}}-e^{|\omega|}
-e^{-\delta|\omega|}\left(-1+e^{|\omega|}\right)
+2N_{q}\cos\left(b\omega\right)}
{2\left(1+2\cosh\left(|\omega|\right)\right)}.
\end{align}
It follows that 
\[
M_{1}=\int d\lambda\tilde{\rho}_{1}(\lambda)=\frac{2N_{q}}{3},
\qquad
M_{2}=\int d\mu\tilde{\rho}_{2}(\mu)=\frac{N_{q}}{3}.
\]
Hence the Dynkin labels of this state are $(1,0)$. This single fundamental quark excitation arises because the impurity quark remains unscreened.

We can add the boundary string solution of rank-1 equation $\lambda_\alpha=i(\frac{1}{2}-\delta$ and the resultant boundary string solution $\mu_k=i(1-\delta)$ of the rank-2 equation, we obtain the new Bethe Ansatz equations of the form
\begin{align}
    2\tilde{\rho}_{1}(\omega)&=\tilde{a}_{\frac{1}{2}}(\omega)(1+2N_q\cos(\omega b))+ 2\tilde{K}^{\{2\}}(\omega)\tilde{\rho}_2(\omega)-2\tilde{K}^{\{1\}}(\omega)\tilde{\rho}_1(\omega)-1\\
     2\tilde{\rho}_{2}(\omega)&=\tilde{a}_{\frac{1}{2}}(\omega)+2\tilde{K}^{\{2\}}(\omega)\tilde{\rho}_1(\omega)-2\tilde{K}^{\{1\}}(\omega)\tilde{\rho}_2(\omega)-1+e^{-((\delta-2 ) | \omega | )}-e^{-((\delta -1) | \omega | )}.
\end{align}
Solving these coupled equations, we find
\begin{align}
    \tilde\rho_1(\omega)&=\frac{4 N_q \cosh \left(\frac{| \omega | }{2}\right) \cos (b \omega )+\left(e^{| \omega | }-1\right) e^{\frac{1}{2} (3-2 \delta ) |
   \omega | }+e^{-\frac{| \omega | }{2}}-e^{| \omega | }}{2 (2 \cosh (| \omega | )+1)}\\
   \tilde\rho_2(\omega)&=\frac{\left(e^{2 | \omega | }-1\right) e^{| \omega | -\delta  | \omega | }+e^{-\frac{| \omega | }{2}}-e^{| \omega | }+2 N_q \cos (b
   \omega )}{2 (2 \cosh (| \omega | )+1)}
\end{align}

The total number of rank-1 roots including the one purely imaginary boundary string solution is $M_1=1+\int\rho_1(\lambda)\mathrm{d}\lambda=\frac{1}{3} \left(2 N_q+3\right)$ and the total number of rank-2 roots including the one purely imaginary boundary string solution is $M_2=1+\int\rho_2(\mu)\mathrm{d}\mu=\frac{1}{3} \left(N_q+3\right)$. Since the Dynkin labels $(p,q)=(0,-1)$, this is not a valid state as one of the Dynkin labels is negative. Thus, to make it a valid state, we need to add a hole in the rank 2 equations at position $\vartheta$, such that the new Bethe Ansatz equations become
\begin{align}
    2\tilde{\rho}_{1}(\omega)&=\tilde{a}_{\frac{1}{2}}(\omega)(1+2N_q\cos(\omega b))+ 2\tilde{K}^{\{2\}}(\omega)\tilde{\rho}_2(\omega)-2\tilde{K}^{\{1\}}(\omega)\tilde{\rho}_1(\omega)-1\\
     2\tilde{\rho}_{2}(\omega)&=\tilde{a}_{\frac{1}{2}}(\omega)+2\tilde{K}^{\{2\}}(\omega)\tilde{\rho}_1(\omega)-2\tilde{K}^{\{1\}}(\omega)\tilde{\rho}_2(\omega)-1+e^{-((\delta-2 ) | \omega | )}-e^{-((\delta -1) | \omega | )}-2\cos(\omega\vartheta).
\end{align}
and the solution of the coupled equations becomes
\begin{align}
    \tilde\rho_1(\omega)&=\frac{e^{-\frac{| \omega | }{2}}-e^{| \omega | }+e^{\frac{1}{2} (3-2 \delta ) | \omega | } \left(-1+e^{| \omega | }\right)-2 e^{\frac{|
   \omega | }{2}} \cos ( \omega \vartheta )+4 \cos (b \omega ) \cosh \left(\frac{| \omega | }{2}\right) N_q}{2 (1+2 \cosh (| \omega | ))}\\
   \tilde\rho_2(\omega)&=\frac{e^{-\frac{| \omega | }{2}}-e^{| \omega | }+e^{| \omega | -\delta  | \omega | } \left(-1+e^{2 | \omega | }\right)-2 \cos (\vartheta 
   \omega )-2 e^{| \omega | } \cos ( \omega \vartheta )+2 \cos (b \omega ) N_q}{2 (1+2 \cosh (| \omega | ))}.
\end{align}

The total number of rank-1 roots including the one purely imaginary boundary string solution is $M_1=1+\int\rho_1(\lambda)\mathrm{d}\lambda=\frac{2}{3} \left(N_q+1\right)$ and the total number of rank-2 roots including the one purely imaginary boundary string solution is $M_2=1+\int\rho_2(\mu)\mathrm{d}\mu=\frac{1}{3} \left(N_q+1\right)$. The Dynkin labels of this state os $(p,q)=(0,0)$. Hence, it is a triality  1 state where the unscreened quark and the bulk itinerant antiquark form a meson singlet state. 

The change in root density of the rank 1 equation due to the presence of the two boundary string solutions and the hole in the rank 2 equation is
\begin{equation}
    \Delta\rho_1(\omega)=-\frac{1}{2} \left(1-e^{| \omega | }\right) e^{\frac{1}{2} (1-2 \delta ) | \omega | }-\frac{e^{\frac{| \omega | }{2}} \cos ( \omega \vartheta )}{2 \cosh (| \omega | )+1}.
\end{equation}

The first term is the contribution from the two boundary string solutions, and the second term is the contribution from the hole in the rank-2 equation. The energy of the hole in the rank-2 equation in the scaling limit is $E_\vartheta=m\cosh\left(\frac{2\pi}{3}\vartheta\right)$, which takes the minimum value of $m$ as $\vartheta\to\infty$. 

The energy of the two boundary string solutions becomes
\begin{align}
    E_\delta=-\tan ^{-1}\left(\frac{b}{b^2+(\delta -1) \delta }\right)+\frac{D}{2\pi}\int \left(-\frac{1}{2} \left(1-e^{| \omega | }\right) e^{\frac{1}{2} (1-2 \delta ) | \omega | } \right)\left(\frac{2i\pi}{\omega}e^{-\frac12|\omega|-ib\omega} \right) \mathrm{d}\omega=0
\end{align}
where the first term represents the bare energy of the rank-1 boundary string solution, while the second term accounts for the energy shift arising from both boundary strings through the modification of the rank-1 root density. In the thermodynamic limit, these two contributions exactly cancel, causing the combined energy of the two boundary strings to vanish.

All other excitations are obtained by introducing suitable numbers of holes of each rank, together with the corresponding bulk string solutions, so that the resulting states have both Dynkin labels positive. There are no such low-lying excitations in which either a single particle bound mode or a multiparticle Kondo effect screens the impurity quark in this regime. 
\section{Renormalization group analysis}\label{RGeqn-deriv}

Recalling the relation between the physical mass of a quark in terms of the cut-off $D$ and running coupling constant $b$
\begin{equation}
    \bar m = \frac {4 D} {\pi} e^{-\frac {2\pi b} {3}},
\end{equation}
we invert the relation to write the bulk running coupling constant as
\begin{equation}
    b(D)=\frac{3 \log \left(\frac{4 D}{\pi  \bar m}\right)}{2 \pi }.
\end{equation}
Now, using $b=\frac {9 - 32 g^2} {72 g}$, and for small $g$ writing $b\approx \frac {1} {8 g}$, we write the coupling constant $g$ as a function of the cutoff $D$ as
\begin{equation}
    g(D)=\frac {1} {12\log\left(\frac {4 D} {\pi \bar m} \right)}.
    \label{gdrel}
\end{equation}

Differentiating Eq.~\eqref{gdrel} with respect to the log of the cutoff $D$, we obtain
\begin{equation}
    \beta(g)=\frac{\mathrm{d}}{\mathrm{d}\ln D}g=-\frac {12 g^2} {\pi}.
\end{equation}

Moreover, from the relation between the running coupling constants $c$ and $b$ and the RG invariant $d$
\begin{equation}
  b^2 - \frac {2 b} {c} - 1 = d^2,
\end{equation}
we find
\begin{equation}
    d^2=-\frac{\left(9-32 g^2\right) \left(1-\frac{32
   J^2}{9}\right)}{144 g J}+\frac{\left(9-32
   g^2\right)^2}{5184 g^2}-1.
\end{equation}
Noticing that $d$ is RG invariant, we can differentiate both sides of the equation with $\ln D$ to obtain {
    \footnote{The full expression for the $\beta$–function of the boundary coupling $J$, Eq.~\ref{fullJRG}, is derived exactly within the
Bethe–Ansatz cutoff scheme and matches the expression obtained via perturbative
calculations in the universal regime. The pole at $32g^{2} = 9$ is outside the universal regime,
similar to the pole in the Bethe–Ansatz coupling 
$c = 2J / (1 - J^{2})$ in the standard Kondo
model (see Ref.~\cite{andrei1983solution} for a detailed discussion about the difference between the two cut-off schemes). In other words, the reliable domain of this expression is
limited to $g^{2} \ll 1$ and $J^{2} \ll 1$, where it correctly reproduces the weak–coupling flow, while the exact Bethe–Ansatz solution reveals a smooth and analytic crossover from the
UV to the IR fixed point with no actual singularities.
}}
\begin{equation}
    \beta(J)=\frac{\mathrm{d}}{\mathrm{d}\ln D}J=-\frac{6 J}{\pi } \frac{\left(32 g^2+9\right) \left(32 g^2
   J-64 g J^2+18 g-9 J\right)}{\left(32 g^2-9\right)
   \left(32 J^2+9\right)}
   \label{fullJRG}
\end{equation}

For smaller values of couplings $J\ll 1$ and $g\ll 1$, the last relation can be simplified to
\begin{equation}
    \beta(J)\approx -\frac{6J}{\pi}(J-2g).
\end{equation}
These are the weak-coupling RG equations mentioned in the main text.

\end{document}